\documentclass{IET}
\usepackage{longtable}
\addtocounter{table}{-1}

\begin{document}
\title{Thyristor Voltage Equalizing Network for Crowbar Application\\\normalfont''This paper is a preprint of a paper accepted by IET Power Electronics and is subject to Institution of Engineering and Technology Copyright. When the final version is published, the copy of record will be available at IET Digital Library''\thanks{''This paper is a preprint of a paper accepted by IET Power Electronics and is subject to Institution of Engineering and Technology Copyright. When the final version is published, the copy of record will be available at IET Digital Library''}}

\author[1**]{Subhash Joshi T. G.}
\affil{Power Electronics Group, Centre for Development of Advanced Computing,

\noindent Thiruvananthapuram, India}

\author{Vinod John}
\affil{Department of Electrical Engineering, Indian Institute of Science, Bangalore, India

\noindent Email:vjohn@ee.iisc.ernet.in}

\affil[**]{subhashj@cdac.in}

\abstract{Many high voltage applications are realized with series connected thyristors. Voltage imbalance among series connected thyristors during steady state as well as in transients is one of the major concerns. This voltage imbalance is mitigated by using static and dynamic balancing network. Dynamic balancing networks are typically designed based on reverse recovery charge of the thyristor during turn-off, which suits many applications. But this is not the case for a crowbar application, where turn-off of the thyristor is not a major circuit constraint. This paper proposes the design method for dynamic balancing network considering gate turn-on delay time and the balancing network component tolerances. The paper derives two models for the dynamic balancing network based on its charge-discharge cycle. The importance of charge-discharge cycle in the design of dynamic balancing network during high $di/dt$ operation is emphasized. Influence of dynamic balancing resistance and crowbar current limiting inductance on voltage imbalance, charging current and discharging current is studied using the analytical model. The proposed design method also offers flexibility to incorporate differences in propagation delays among the thyristor drivers that are used to trigger individual thyristors. Such delays cannot be directly incorporated in the conventional balancing network design method based on reverse recovery. Further, it is also analytically shown that designing the dynamic balancing network based on reverse recovery charge makes the balancing network lossy and bulky for crowbar application. Simulation studies and experimental results on a $12kV$, $1kA$ crowbar consisting of six series connected thyristors confirms the theoretical analysis and validates the proposed approach for crowbar applications.}

\maketitle
\section*{Nomenclature}
\addcontentsline{toc}{section}{Nomenclature}
\begin{longtable}{ll}
$R_s$, $a_R$ & Static balancing resistance and its tolerance.\\
$R_d,C_d$ & Dynamic balancing resistance and capacitance.\\
$a_c$& Tolerance of dynamic balancing capacitor.\\
$L$&$di/dt$ limiting inductor of crowbar.\\
$N$& Number of thyristors connected in series.\\
$v_G,i_G$ & Thyristor gate to cathode voltage and gate current.\\
$v_{AK},i_{A}$& Thyristor anode to cathode voltage and anode current.\\
$v_{AK,N}$& $N^{th}$ thyristor anode to cathode voltage.\\
$V_{AK,1\_max}$& Maximum voltage across thyristor $T_1$ during triggering of $T_2$ to $T_N$.\\
$v_{C_d,L}$& Voltage across the first thyristor and $L$.\\
$V_s$& Operating dc voltage of crowbar.\\
$t_{dmax},t_{dmin}$ & Maximum and minimum turn on delay time.\\
$t_{dTol}$&$t_{dmax}-t_{dmin}$\\
$t_{on}$& Time taken by $v_{AK}$ to drop from $100\%$ of forward blocking voltage to $0V$.\\
$i_{ch},i_{dis}$ & Charging and discharging current of dynamic balancing network.\\
$I_{ch\_max}$, $I_{dis\_max}$& Maximum charging and discharging current of dynamic balancing network.
\end{longtable}

\section{Introduction}
An increasing number of power electronic systems are being used in high voltage applications, such as high power drives, high voltage un-interruptable power supply, pulse power systems, static VAR compensator and crowbar switches~\cite{IEEEhowto:Jensen}. Among the various semiconductor devices available for design, thyristor is widely used in various power electronic systems when the operating voltages are in the range of kilovolts. Availability of thyristors at higher voltage and current rating makes thyristor a good choice for such applications. In many high voltage applications, meeting the required voltage rating with single thyristor is not feasible. Hence, there is a need for series connection of the thyristor switches~\cite{IEEEhowto:Tooker}.

\begin{figure}[!t]
\centering
\subfigure[]{\includegraphics[keepaspectratio,scale=1.00]{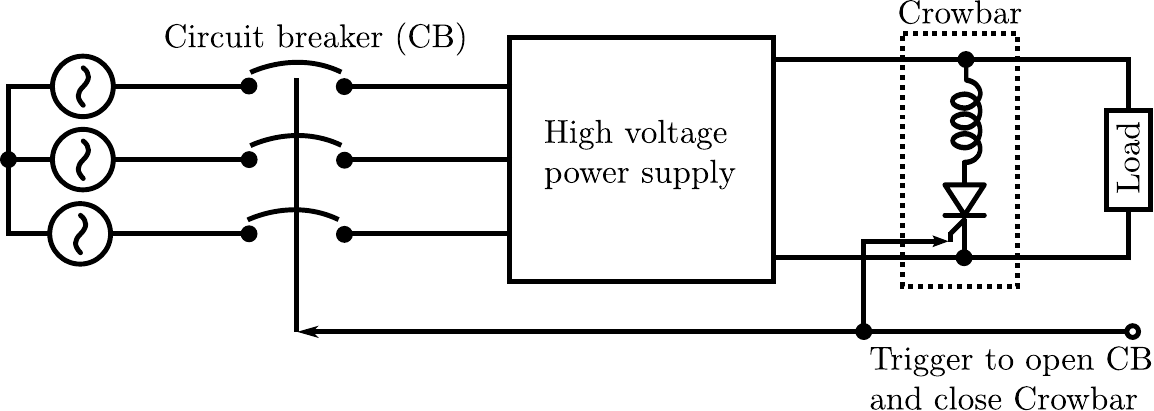}}\\
\subfigure[]{\includegraphics[keepaspectratio,scale=1.00]{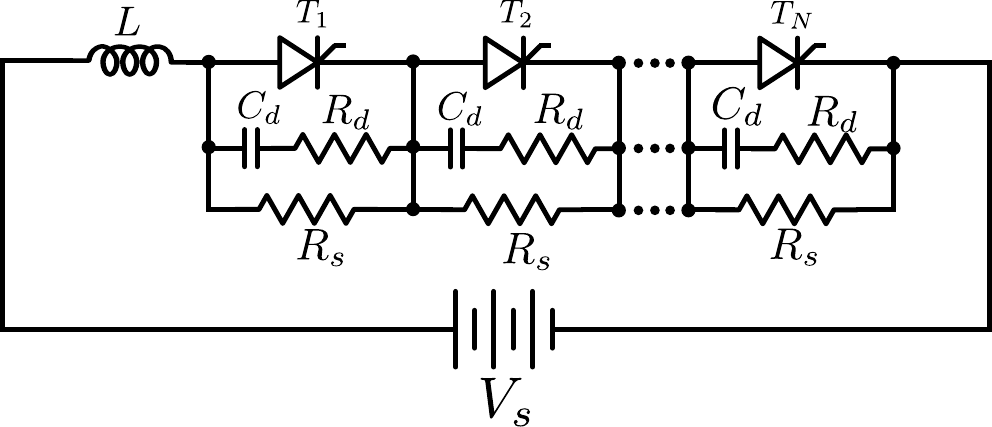}}\\
\subfigure[]{\includegraphics[keepaspectratio,scale=1.00]{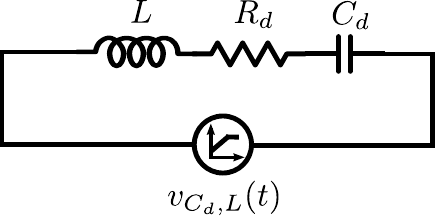}}\hspace{0.5in}
\subfigure[]{\includegraphics[keepaspectratio,scale=1.00]{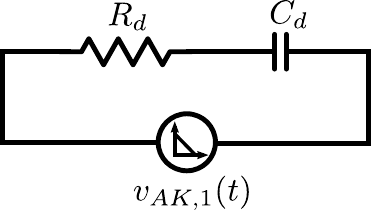}}
\caption{Schematic of crowbar circuit showing
\subcaption{a.}{Power supply and load connection at dc output.}
\subcaption{b.}{Static and dynamic balancing network.}
\subcaption{c.}{Simplified equivalent circuit when thyristor $T_1$ is off and other $(N-1)$ thyristors are triggered on initially.}
\subcaption{d.}{Simplified equivalent circuit when thyristor $T_1$ is turned on subsequently.}}
\label{fig_crowbar}
\end{figure}

A crowbar is a fault energy-diverting element built with thyristors, connected at the output of high voltage dc source as shown in Fig.~\ref{fig_crowbar}(a). The dc source  feeds power to sensitive loads, like microwave or plasma tubes. The triggering of crowbar is initiated by turning on the thyristors when a fault signal is received from fault sensing circuits in the load. This can be for conditions of over voltage, or short circuit, or any other situations which activate the protection circuits. When the crowbar receives the trigger signal from a protection circuit, all the series connected thyristors are turned-on and fault energy, from the storage devices such as capacitors and the follow-on current from the input power supply, is diverted through crowbar within a few microseconds. Concurrent to this, the trigger signal is also transmitted to open the input circuit breaker (CB). The crowbar is kept ON until the input CB opens, which can take as long as $100ms$. This along with the triggering of the crowbar reduces the dc voltage close to zero~\cite{IEEEhowto:Srinivas}.

\begin{figure}[!t]
\centering
\subfigure[]{\includegraphics[keepaspectratio,scale=1.00]{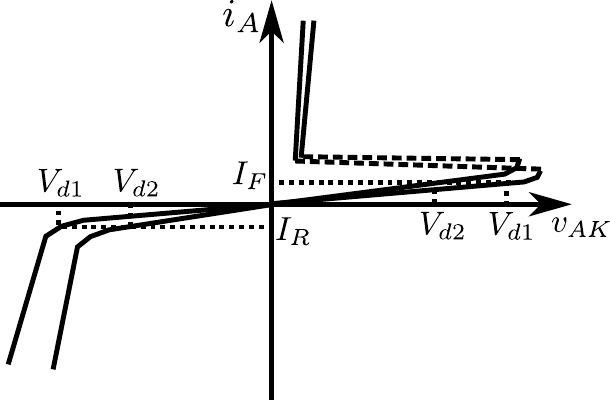}}\hspace{0.5in}
\subfigure[]{\includegraphics[keepaspectratio,scale=1.00]{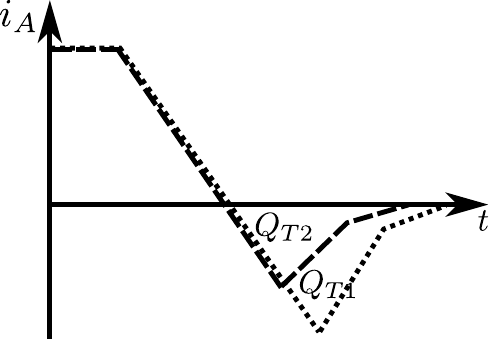}}\\
\subfigure[]{\includegraphics[keepaspectratio,scale=.5]{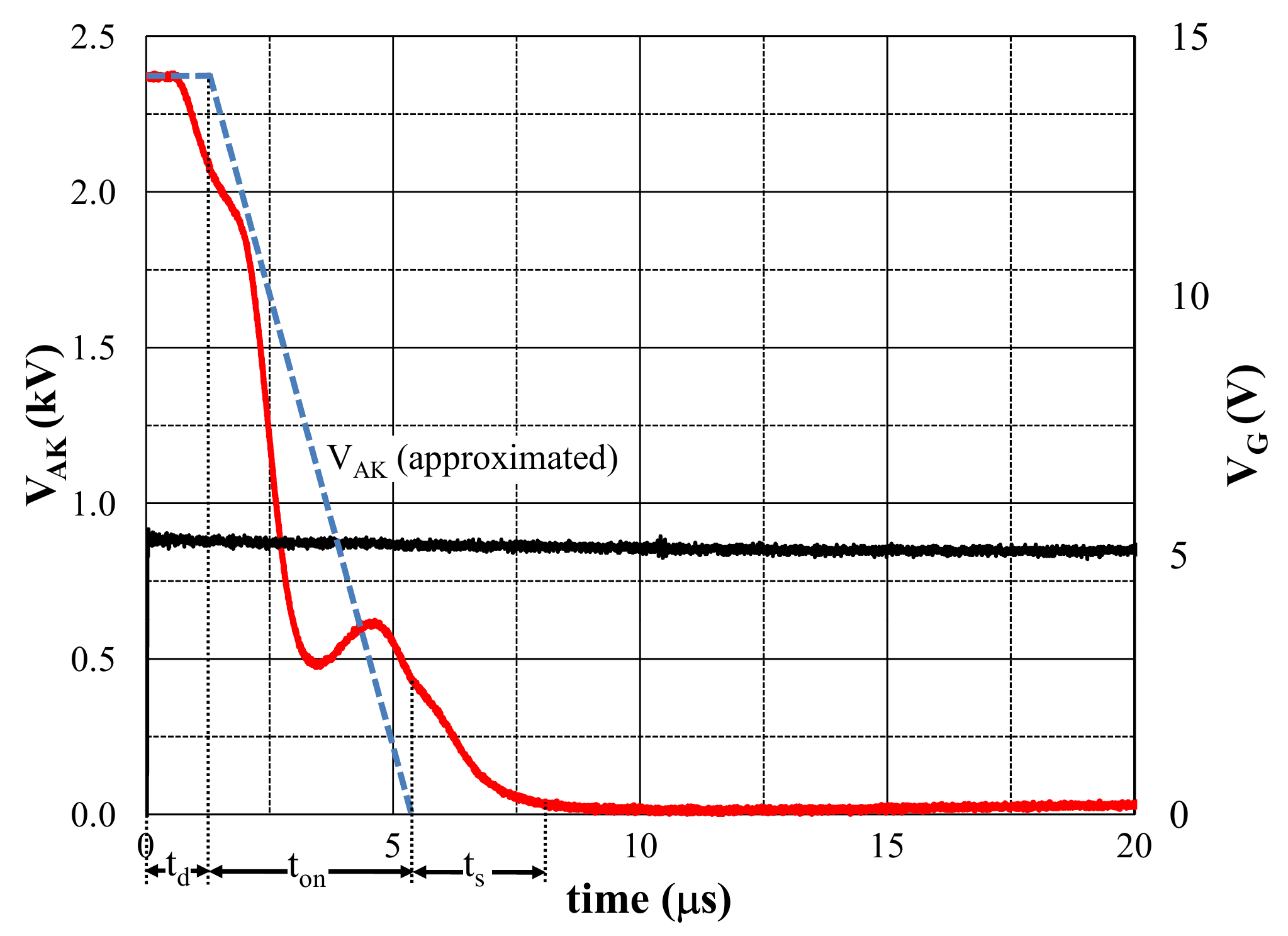}}
\caption{Characteristics showing
\subcaption{a.}{Mismatch in two thyristor V-I characteristics.}
\subcaption{b.}{Mismatch in two thyristor reverse recovery characteristics.}
\subcaption{c.}{$v_{AK}$ and $v_G$ waveforms showing turn-on characteristics of thyristor measured in the laboratory.}}
\label{fig_chara}
\end{figure}

While connecting thyristors in series, sharing of voltage by each thyristor during steady state as well as in dynamic condition is one of the major concerns to be addressed~\cite{IEEEhowto:Waldmeyer}. The main cause of voltage imbalance in series connected thyristors during steady state is due to their difference in reverse blocking leakage current. If the \textit{V-I} characteristics of thyristors connected in series are different, for the same leakage current, the voltage sharing among thyristors will be unequal as illustrated in Fig.~\ref{fig_chara}(a). Passive resistors $R_s$ called static balancing networks are connected across each thyristor as shown in Fig.~\ref{fig_crowbar}(b) as a solution for static voltage imbalance. The value of $R_s$ is selected based on the reverse leakage current of thyristor~\cite{IEEEhowto:link1}.


When series connected thyristors are turned-off transient voltage imbalances can occur among them due to mismatch in their reverse recovery charge~\cite{IEEEhowto:Nancy}. When the thyristor is turned-off, the charge carriers stored in the device need to be removed completely before it recovers its voltage blocking capability~\cite{IEEEhowto:Qiang,IEEEhowto:Taib}. The difference in the amount of charge stored in the junction makes the turn off time to vary from one thyristor to another as illustrated in Fig.~\ref{fig_chara}(b). Passive resistor-capacitor network ($R_d,C_d$), shown in Fig.~\ref{fig_crowbar}(b), called dynamic balancing network, is used as a solution for transient voltage imbalance. These are designed based on the mismatch in reverse recovery charge of the thyristors~\cite{IEEEhowto:Zhankai}.

Conventionally the static and dynamic balancing networks are designed by considering two modes of operation of thyristors~\cite{IEEEhowto:Rivet,IEEEhowto:Backlund}; in mode 1, thyristors are in blocking stage and in mode 2, thyristors are being turned-off. These modes of operation of the thyristors reflect actual operating conditions in most applications. Hence in many literature the balancing network are designed considering turn-off condition also~\cite{luji,abba,bai}. But this is not true in case of crowbar operation, where in mode 1, thyristors are in forward blocking stage and in mode 2, thyristors are being turned-on. This is shown to require new constraints for selection of the static and dynamic balancing network components. In~\cite{zakar}, switching-on over voltage across series connected thyristor is discussed, but the over voltage addressed is due to the firing capacitor connected across gate and cathode. For a three-phase bridge circuit, the turn-on over voltage during commutation of the complementary switch is discussed in~\cite{karad}, but the over voltage is solved numerically only.

When series connected thyristors are turned-on, voltage imbalance can occur due to difference in gate turn-on delay time. In~\cite{ref16}, the design of a dynamic balancing network based on gate turn-on delay time is addressed. The analysis detailed in~\cite{ref16} did not consider the influence of dynamic balancing resistance and the crowbar current limiting inductance. In this paper the analysis is extended by considering the influence of dynamic balancing resistance. The paper derives two models for the dynamic balancing network based on its charge-discharge cycle. The importance of charge-discharge cycle in the design of dynamic balancing network during high $di/dt$ operation of crowbar is also emphasized in the paper. Influence of dynamic balancing resistance and crowbar current limiting inductance on voltage imbalance, charging current and discharging current is also explored. The analysis shows that for high $di/dt$ applications the charging current of dynamic balancing network can be higher than the discharging current. Hence by considering only the discharging current as done in the conventional method, the design leads to large deviations in the results. The paper compares this design with that using traditional dynamic balancing component design based on reverse recovery charge. It is also found that the proposed design provides flexibility to include the difference in propagation delays among the gate driver circuits used in triggering individual thyristors. This allows for a design without complex pulse synchronizing circuit used in crowbar applications. Tolerances of components are also considered in the proposed balancing network design. The analytical design results are validated using time domain simulations. Experimental results using a laboratory crowbar prototype confirms the theoretical analysis.
\section{Thyristor Characteristics}
Turn-on time of thyristor can be divided into three parts (a) delay time (b) rise time and (c) conduction spreading time~\cite{IEEEhowto:Bergman}. After initiating the gate-to-cathode current $I_G$, an appreciable time $t_d$ is required to establish the charge in thyristor to support a current greater than its holding current. During rise time, $i_A$ rises rapidly in a small area near the vicinity of the gate and a similar decrease in voltage occurs between anode and cathode $v_{AK}$. The area of conduction spreads during spreading time until the whole cathode starts conducting. In many applications, evaluating the delays of $i_A$ rise and $v_{AK}$ fall with respect to gate voltage signal is better suited than with respect to gate current signal. Also, while using devices such as Light Triggered Thyristor (LTT) the gate currents are not accessible. In the datasheet $t_d$ is defined as the time from start of the triggering pulse $v_G$ to $v_{AK}$ dropping below 90\% of the applied forward off-state voltage~\cite{IEEEhowto:Melanie}, as shown in a sample experimental waveform given in Fig.~\ref{fig_chara}(c). The time during which $v_{AK}$ drops from $90\%$ to $20\%$ is referred as turn on time, where as during spread time $t_s$, $v_{AK}$ drops from $20\%$ to forward on voltage. However for the analysis the variation of $v_{AK}$ is assumed to be linear as shown by dotted line in Fig.~\ref{fig_chara}(c). Such an approximation shows a worst case $v_{AK}$, since the approximated $v_{AK}$ is delayed more than the actual curve. In the approximated $v_{AK}$, turn on time $t_{on}$ is redefined as the time taken by $v_{AK}$ to drop from $100\%$ of forward blocking voltage to $0V$. Thyristors having smaller $t_{d}$ will turn-on earlier than the one having higher $t_{d}$. This leads to voltage imbalance among series connected thyristors. Hence dynamic balancing network consisting of $R_d$ and $C_d$, should be designed based on the gate turn-on delay time mismatch of the thyristors used. From the datasheet of thyristors~\cite{IEEEhowto:abb}, it is found that forward leakage current as well as reverse leakage current are similar. Hence the method used to design the static balancing network $R_s$ remains similar to the conventional method. However in the proposed method the worst case balancing network component tolerances is also considered in the design procedure.
\section{Proposed design of static and dynamic balancing network}
Inductor $L$ connected in series with the thyristor shown in Fig.~\ref{fig_crowbar}(b), ensures that the slope of the current does not cause any damage to the thyristor. The upper limit of the value of this inductor is based on the amount of fault energy that can be tolerated by the load of the crowbar shown in Fig.~\ref{fig_crowbar}(a). The lower limit of the value of the inductor is based on the $di/dt$ rating of the thyristor.
\subsection{Charging cycle of dynamic balancing network}
Let $N$ thyristors connected in series be fed from a dc source, $V_s$, through a $di/dt$ limiting inductor, $L$, as shown in Fig.~\ref{fig_crowbar}(b). For the dynamic balancing network analysis the following assumptions are made:
\begin{enumerate}
\item For worst case analysis the thyristor $T_1$ is chosen to have maximum gate turn-on delay time $t_{dmax}$ and remaining $(N-1)$ thyristors have minimum turn-on delay time $t_{dmin}$. The difference between the maximum and minimum thyristor turn-on delay time is defined as,
\begin{equation}
t_{dTol}=t_{dmax}-t_{dmin}
\end{equation}
\item During the turn-on process of thyristor, anode-to-cathode voltage varies linearly and the curve is identical for all thyristors except it is delayed by its own gate turn-on delay time.
\item Influence of the static balancing network $R_s$ on dynamic balancing network is assumed to be minimal due to its relative high impedance.
\item  For the worst case analysis the capacitor connected across the first thyristor is chosen to have the lowest tolerance limit $a_c$ where $a_c$ is defined as $C_d\in \{C_{d,nominal}(1\pm a_c)\}$
\end{enumerate}

The voltage appearing across each thyristor during forward blocking state is $V_s/N$. $t_{on}$ is the time taken by $v_{AK}$ to reach zero from its forward blocking voltage, $V_s/N$. Then the expression for the variation of voltage across $(N-1)$ thyristors at time $t$, in the time duration $t_{dmin}\le t\le (t_{on}+t_{dmin})$ is given by,
\begin{align}\label{vd}
v_{AK,2}(t)=v_{AK,3}(t)=\dots=v_{AK,N}(t)=-\frac{V_s}{Nt_{on}}(t-(t_{on}+t_{dmin}))
\end{align}
Total voltage appearing across $(N-1)$ thyristor is,
\begin{align}
v_{\left[AK,2~to~N\right]}(t)=-\frac{(N-1)V_s}{Nt_{on}}(t-(t_{on}+t_{dmin}))
\end{align}
Since the voltage transition across the first device $T_1$ is not yet initiated, the dynamic balancing network connected across $T_1$ will be in charging cycle. During this cycle the voltage appearing across $L$ and $T_1$, which includes the balancing network, is shown by a simplified equivalent circuit in Fig.~\ref{fig_crowbar}(c) and is given by,
\begin{align}
v_{C_d,L}(t)&=V_s-V_{\left[AK,2~to~N\right]}(t)\nonumber\\
v_{C_d,L}(t)&=\frac{V_s}{N}+\frac{V_s(N-1)(t-t_{dmin})}{Nt_{on}}\label{vcdL1}
\end{align}
Based on the relative value of $t_{on}$ and $t_{dTol}$ during this cycle, the $v_{C_d,L}(t)$ takes following values
\begin{enumerate}
\item If $t_{dTol}\le t_{on}$
\begin{align}
v_{C_d,L}(t)=\frac{V_s}{N}+\frac{V_s(N-1)(t-t_{dmin})}{Nt_{on}}\label{vcdL}
\end{align}
\item If $t_{dTol}> t_{on}$, the $(N-1)$ thyristors turn on transitions will be completed at $t=(t_{dmin}+t_{on})$. Hence,
\begin{eqnarray}
v_{C_d,L}(t) = \left \{
\begin{aligned}
 &\frac{V_s}{N}+\frac{V_s(N-1)(t-t_{dmin})}{Nt_{on}}, && \text{if}\ t\le(t_{dmin}+t_{on}) \label{vcdL2}\\
 &V_s, && \text{if}\ t>(t_{dmin}+t_{on})
\end{aligned}  \right.
\end{eqnarray} 
\end{enumerate}
Hence the charging cycle of dynamic balancing network include two dynamic models based on relative value of $t_{on}$ and $t_{dTol}$, which are considered below.
\subsubsection{Dynamic model for $t_{dTol}\le t_{on}$}
The simplified equivalent circuit is shown in Fig.~\ref{fig_crowbar}(a) where $v_{C_d,L}(t)$ is given by (\ref{vcdL}). If $i_{ch}(t)$ is the charging current of dynamic balancing network, the dynamic equation related to this circuit is given by,
\begin{align}\label{dyn1}
L\frac{d^2i_{ch}}{dt^2}+R_d\frac{di_{ch}}{dt}+\frac{i_{ch}}{(1-a_c)C_d}=\frac{V_s(N-1)}{Nt_{on}}
\end{align}
The minimum value of tolerance is chosen for $C_d$ across thyristor $T_1$ in (\ref{dyn1}), as this gives the worst case over voltage for $T_1$. In most crowbar balancing circuits, $R_d$ is much smaller than the characteristic impedance leads to complex solutions to (\ref{dyn1}). At $t=t_{dmin}$, choosing the initial conditions as $i_{ch}(t)=0$ and voltage across the inductor is zero, the above dynamic equation can be solved as,
\begin{equation}\label{i1}
i_{ch}(t)=\frac{V_s(N-1)(1-a_c)C_d}{Nt_{on}}\left[1-\frac{e^{-\delta(t-t_{dmin})}}{\omega_d\sqrt{L(1-a_c)C_d}}cos\left(\omega_d\left(t-t_{dmin}\right)-\phi\right)\right]
\end{equation}
where, $\delta=\dfrac{R_d}{2L}$,~$\omega_d=\sqrt{\dfrac{1}{L(1-a_c)C_d}-{\left(\dfrac{R_d}{2L}\right)^2}}$ and $\phi=tan^{-1}\left(\dfrac{\delta}{\omega_d}\right)$.
\vspace{\baselineskip}
\\Therefore, the voltage appearing across the first device is given by,
\begin{align}\label{vd1}
v_{AK,1}(t)=\frac{V_s}{N}+\frac{V_s(N-1)}{Nt_{on}}\left[(t-t_{dmin})-\frac{e^{-\delta(t-t_{dmin})}}{\omega_d}sin\left(\omega_d\left(t-t_{dmin}\right)\right)\right]
\end{align}
The second term in (\ref{vd1}) represents the transient term initiated after triggering $(N-1)$ thyristors. Since the first thyristor $T_1$ get triggered at $t=t_{dmax}$, the maximum value of $i_{ch}(t)$ and $V_{AK,1}(t)$ are obtained by choosing $t=t_{dmax}$ in (\ref{i1}) and (\ref{vd1}) respectively given by,
\begin{align}
I_{ch\_max}&=\frac{V_s(N-1)(1-a_c)C_d}{Nt_{on}}\left[1-\frac{e^{-\delta t_{dTol}}}{\omega_d\sqrt{L(1-a_c)C_d}}cos\left(\omega_dt_{dTol}-\phi\right)\right]\label{i1m1i1}\\
V_{AK,1\_max}&=\frac{V_s}{N}+\frac{V_s(N-1)}{Nt_{on}}\left[t_{dTol}-\frac{e^{-\delta t_{dTol}}}{\omega_d}sin\left(\omega_dt_{dTol}\right)\right]\label{vd1m1i1}
\end{align}
\subsubsection{Dynamic model for $t_{dTol}>t_{on}$}
Let $t_1$ be defined as $\left(t_{on}+t_{dmin}\right)$. For $t\le t_1$, $v_{C_d,L}(t)$ hold the first condition of (\ref{vcdL2}) and the solutions are given in (\ref{i1}) and (\ref{vd1}). For $t>t_1$, $v_{C_d,L}(t)$ is given by the second condition of (\ref{vcdL2}) and the dynamic equation can be expressed as,
\begin{align}\label{dyn1_}
L\frac{d^2i_{ch}}{dt^2}+R_d\frac{di_{ch}}{dt}+\frac{i_{ch}}{(1-a_c)C_d}=0
\end{align}
This can be solved by applying initial conditions (a) $i_{ch}(t)$ at $t=t_1$ from (\ref{i1}) denoted as $I_{ch(t_1)}$ and (b) voltage across $C_d$ at $t_1$ denoted by $V_{c(t_1)}$ obtained as $\left(V_{AK,1(t_1)}-I_{ch(t_1)}R_d\right)$. $V_{AK,1(t_1)}$ is given by $V_{AK,1}(t)$ at $t=t_1$ obtained from (\ref{vd1}). The solution of (\ref{dyn1_}) is,
\begin{equation}\label{i2}
i_{ch}(t)=e^{-\delta(t-t_{1})}\left[K_1cos\left(\omega_d\left(t-t_1\right)\right)+K_2sin\left(\omega_d\left(t-t_1\right)\right)\right]
\end{equation}
where, $K_1=I_{ch(t_1)}$ and $K_2=\dfrac{V_s-V_{c(t_1)}-I_{ch(t_1)}L\delta}{L\omega_d}$.
\vspace{\baselineskip}
\\Therefore, the voltage appearing across the first device is given by,
\begin{equation}\label{dyn1_2}
V_{AK,1}(t)=V_s+Le^{-\delta(t-t_{1})}\biggl[(K_1\omega_d+K_2\delta)sin\left(\omega_d\left(t-t_1\right)\right)+(K_1\delta-K_2\omega_d)cos\left(\omega_d\left(t-t_1\right)\right)\biggr]
\end{equation}
The maximum value of $i_{ch}(t)$ and $V_{AK,1}(t)$ during this mode of operation can be obtained by choosing $t=t_{dmax}$ in (\ref{i2}) and (\ref{dyn1_2}) respectively.
\subsection{Discharging cycle of dynamic balancing network}
When thyristor $T_1$ is turned on at $t=t_{dmax}$, $C_d$ discharges into the thyristor shown by a simplified equivalent circuit given in Fig.~\ref{fig_crowbar}(d). The slope of $v_{AK}$ of first thyristor is assumed to be same as that of $(N-1)$ thyristors. Then the dynamic equation related to Fig.~\ref{fig_crowbar}(d) is given by,
\begin{align}\label{i3}
R_d\frac{di_{dis}}{dt}+\frac{i_{dis}}{(1-a_c)C_d}=-\frac{V_s}{Nt_{on}}
\end{align}
Depending of the type of model used in (\ref{vcdL}) and (\ref{vcdL2}) the initial conditions are chosen. The initial condition for $i_{dis}(t)$, denoted by $I_{dis(t_{dmax})}$, is obtained either from (\ref{i1}) for $t_{dTol}\le t_{on}$ or from (\ref{i2}) for $t_{dTol}>t_{on}$ by choosing $t=t_{dmax}$. Using this initial condition the solution to the dynamic equation (\ref{i3}) is given by,
\begin{align}\label{i3t}
i_{dis}(t)=\left[I_{dis(t_{dmax})}+\frac{V_s(1-a_c)C_d}{Nt_{on}}\right]e^{-\dfrac{(t-t_{dmax})}{R_d(1-a_c)C_d}}-\frac{V_s(1-a_c)C_d}{Nt_{on}}
\end{align}

As time $t$ increases $i_{dis}(t)$ magnitude increases. Hence the maximum value of $i_{dis}(t)$ can be obtained when thyristor $T_1$ is completely turned on. Starting from forward surge voltage, due to turn on of $(N-1)$ thyristor, the time taken by the first thyristor to reach its forward conduction voltage is,
\begin{align}\label{tmax}
t_{on,1}=\frac{V_{AK,1\_max}Nt_{on}}{V_{s}}+t_{dmax}
\end{align}
where $V_{AK,1\_max}$ is obtained either from (\ref{vd1}) for $t_{dTol}\le t_{on}$ or from (\ref{dyn1_2}) for $t_{dTol}>t_{on}$ by choosing $t=t_{dmax}$. The maximum value of $i_{dis}(t)$ can be obtained by choosing $t=t_{on,1}$ in (\ref{i3t}) and is given by,
\begin{align}\label{i3tmax}
I_{dis\_max}=\left[I_{dis(t_{dmax})}+\frac{V_s(1-a_c)C_d}{Nt_{on}}\right]e^{-\dfrac{(t_{on,1}-t_{dmax})}{R_d(1-a_c)C_d}}-\frac{V_s(1-a_c)C_d}{Nt_{on}}
\end{align}
\begin{figure}[!t]
  \centering
\subfigure[]{\includegraphics[keepaspectratio,scale=0.34]{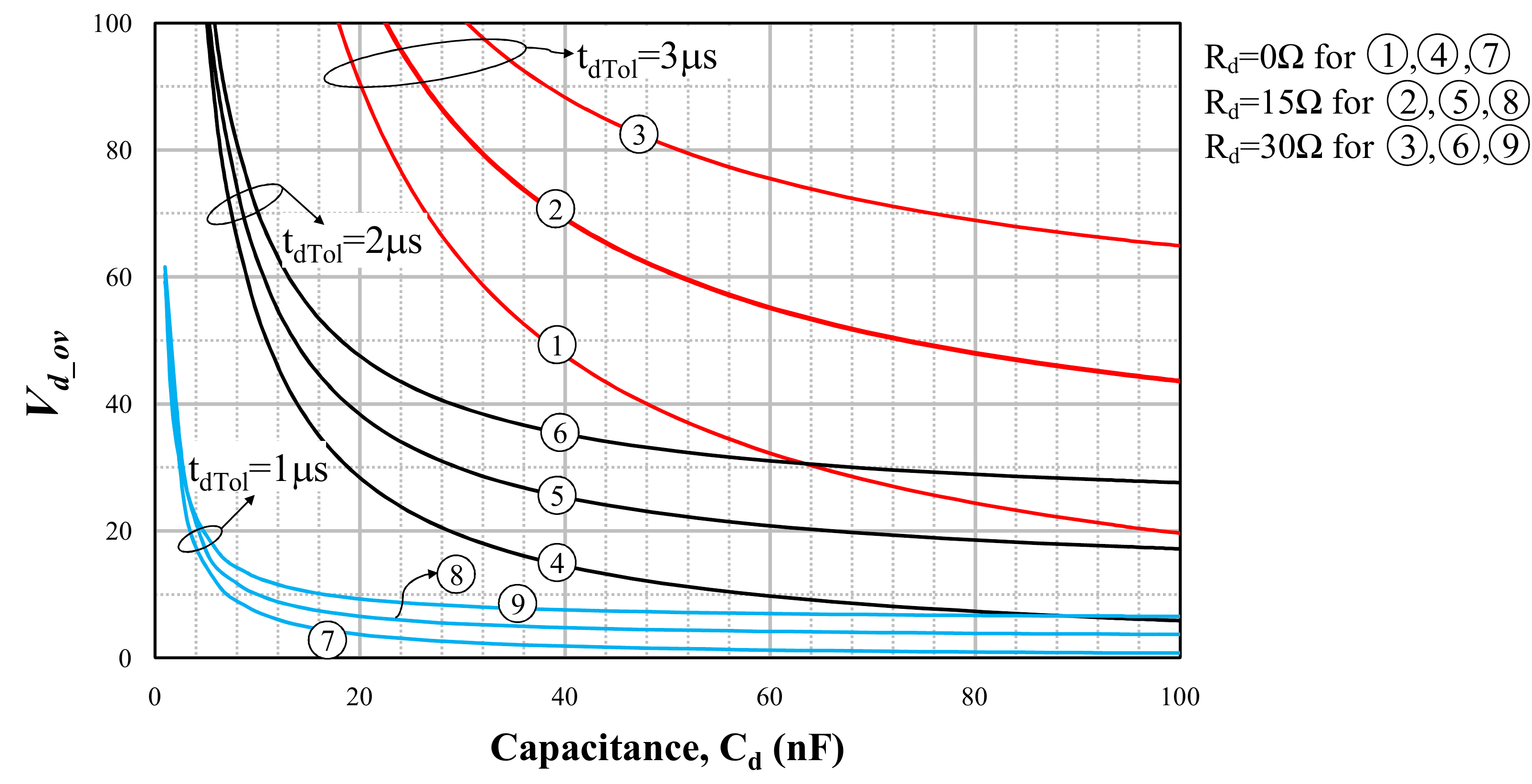}}\\ \hspace{0.05in}
\subfigure[]{\includegraphics[keepaspectratio,scale=0.34]{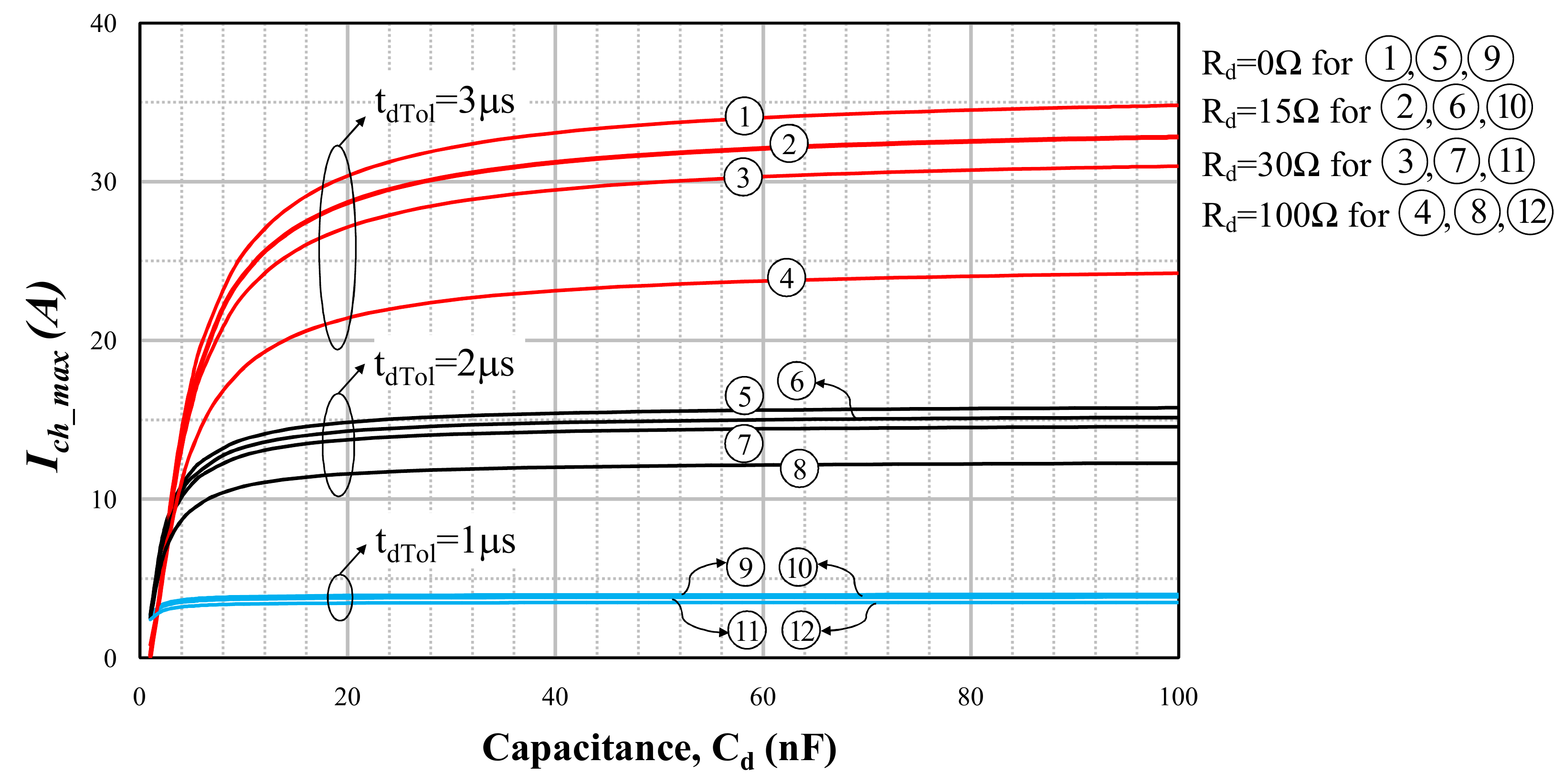}}\\ \hspace{0.3in}
\subfigure[]{\includegraphics[keepaspectratio,scale=0.34]{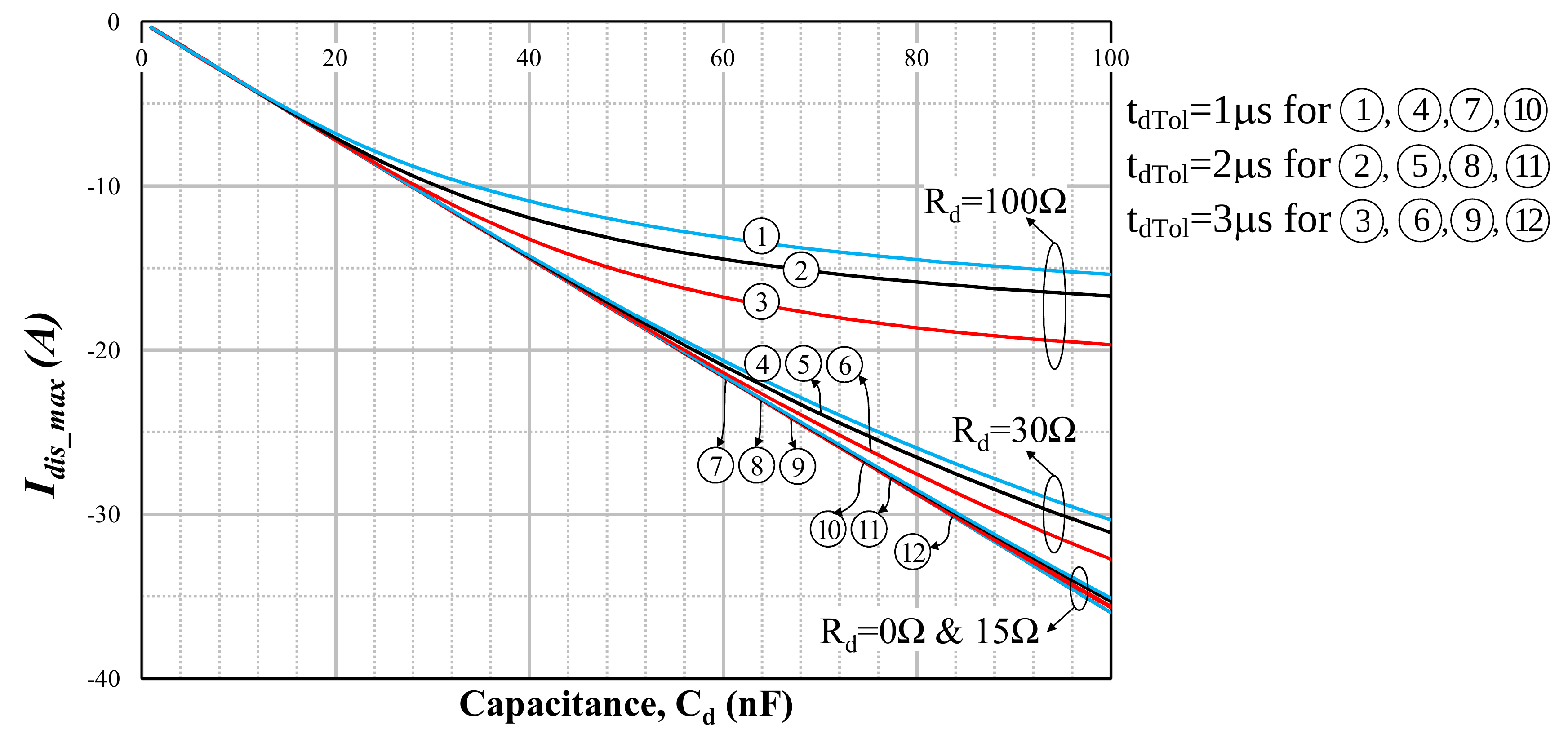}}
\caption{Variation in the thyristor peak stress with $C_d$ for different $t_{dTol}$ and $R_d$, with $t_{on}=5\mu s$
\subcaption{a.}{Percentage of over voltage $V_{d\_ov}$.}
\subcaption{b.}{Maximum charging current $I_{ch\_max}$.}
\subcaption{c.}{Maximum discharging current $I_{dis\_max}$.}}
\label{fig_Vcurve}
\end{figure}
\subsection{Static balancing resistance ($R_s$)}
Design of static balancing network is well established in literature~\cite{IEEEhowto:link1}\cite{IEEEhowto:Rivet}. However, even in this case it is important to consider the tolerance of the resistor $R_s$ used in the balancing network. The static balancing resistance across each thyristor is given by,
\begin{align}\label{Rseq}
R_s=\frac{V_{d1}\left(N(1-a_R)+2a_R\right)-(1+a_R)V_s}{(N-1)(1-a_R^2)(I_{Dmax}-I_{Dmin})}
\end{align}
where, $V_{d1}$ is the maximum allowable steady state forward blocking voltage across thyristor, $I_{Dmax}$ and $I_{Dmin}$ are the maximum and minimum value of forward leakage current and $a_R$ is tolerance limit of selected resistor such that $R_s\in \{R_{s,nominal}(1\pm a_R)\}$.
\section{Analysis of charge-discharge cycle of dynamic balancing network}
\begin{figure}[!t]
  \centering
\subfigure[]{\includegraphics[keepaspectratio,scale=0.30]{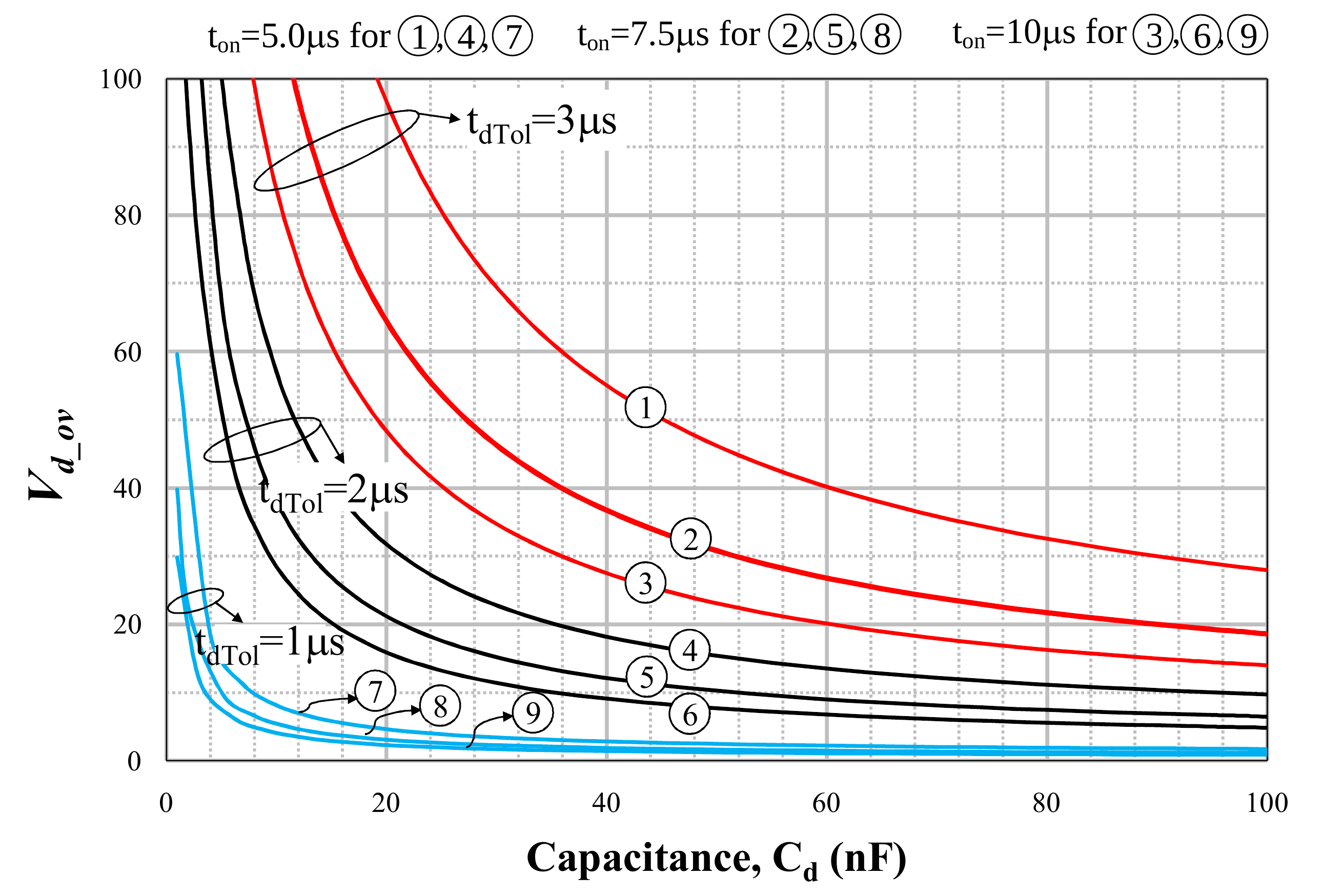}}\hspace{0.05in}
\subfigure[]{\includegraphics[keepaspectratio,scale=0.30]{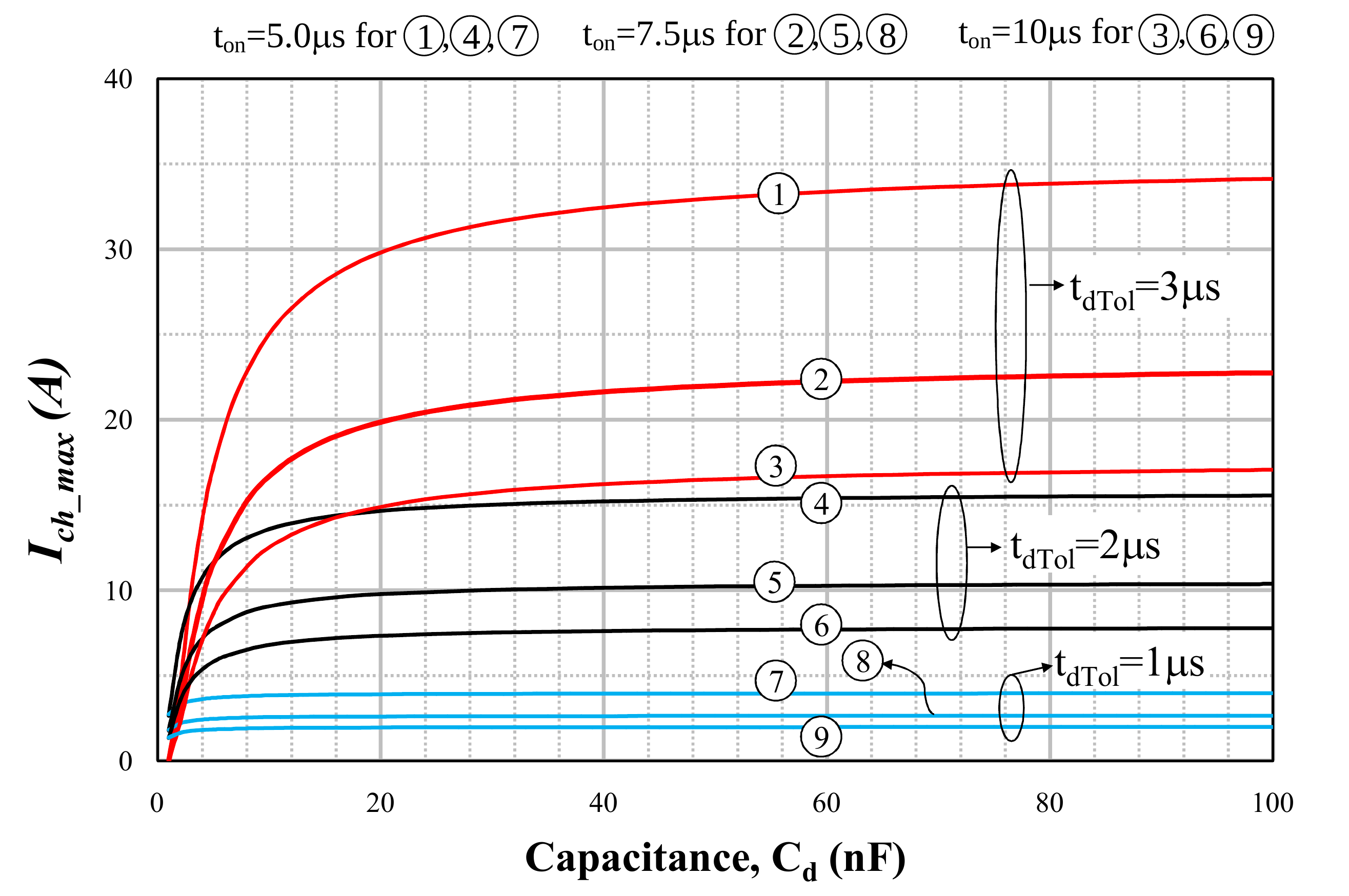}}\hspace{0.05in}
\subfigure[]{\includegraphics[keepaspectratio,scale=0.30]{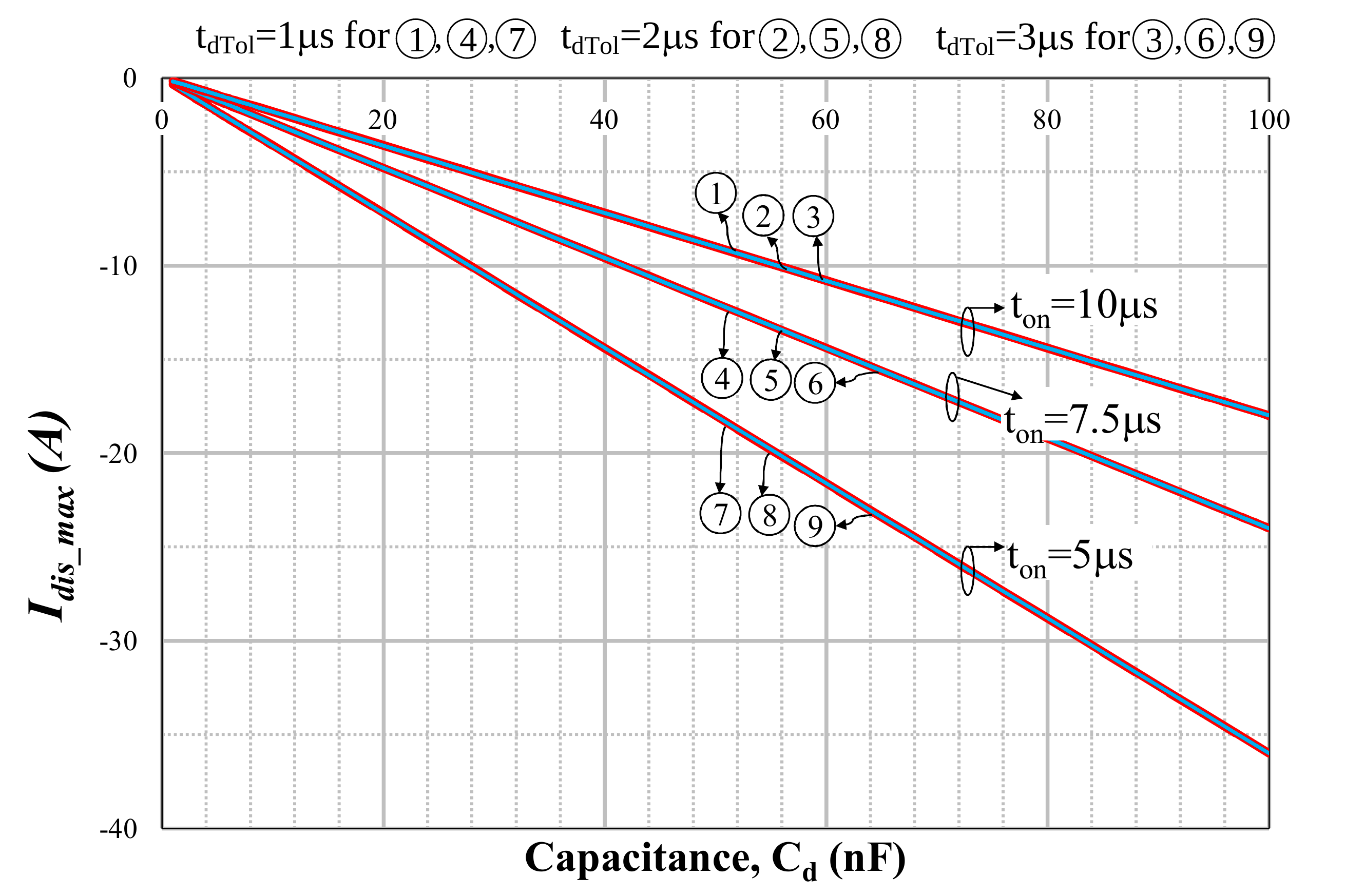}}\hspace{0.05in}
\subfigure[]{\includegraphics[keepaspectratio,scale=0.30]{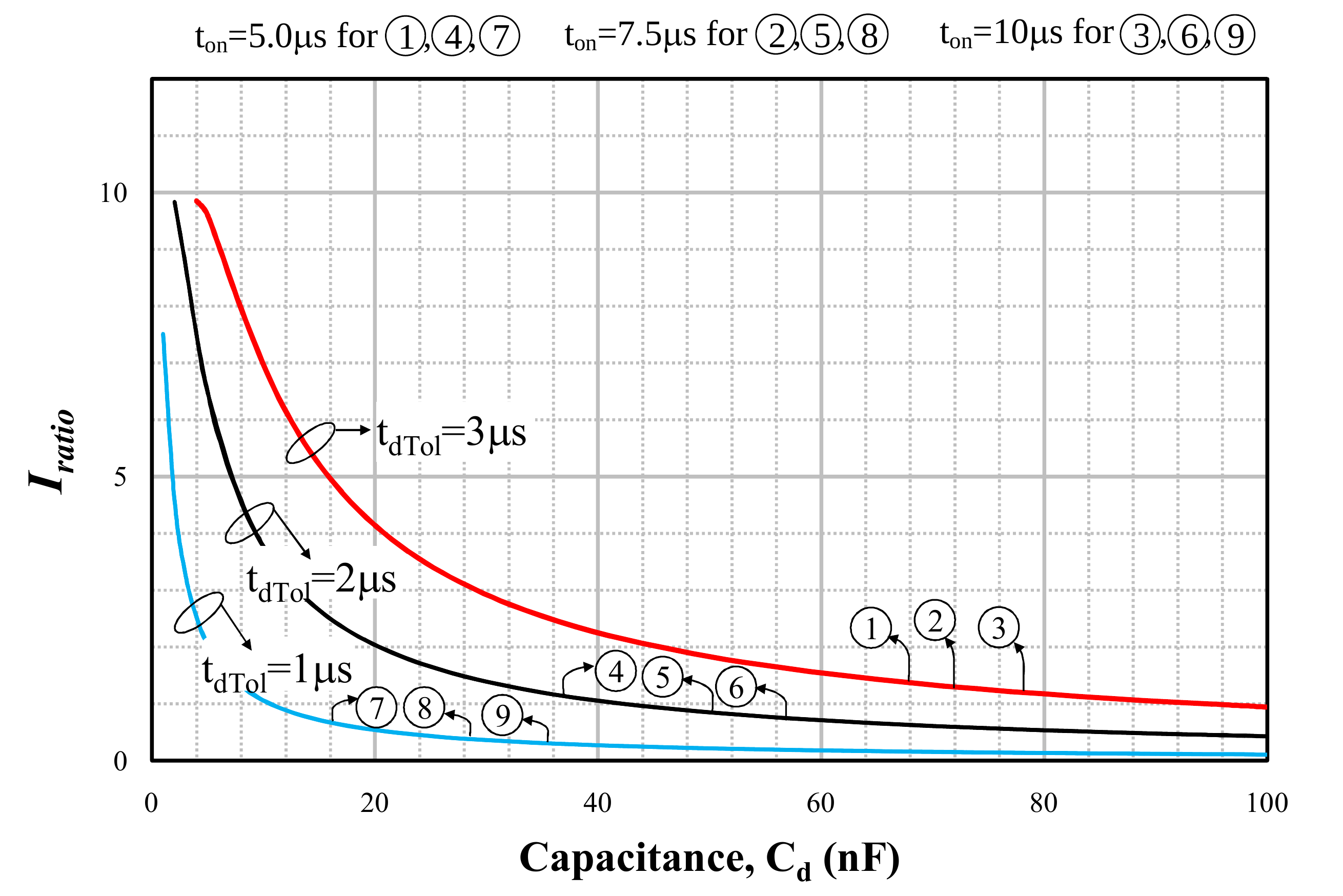}}
\caption{Variation in the following parameters with $C_d$ for different $t_{dTol}$ and $t_{on}$ keeping $R_{d}=5\Omega$
\subcaption{a.}{Percentage of over voltage $V_{d\_ov}$.}
\subcaption{b.}{Maximum charging current $I_{ch\_max}$.}
\subcaption{c.}{Maximum discharging current $I_{dis\_max}$.}
\subcaption{d.}{Ratio of $I_{ch\_max}$ and $I_{dis\_max}$.}}
\label{fig_VIton}
\end{figure}
\begin{figure}[!t]
  \centering
\subfigure[]{\includegraphics[keepaspectratio,scale=0.30]{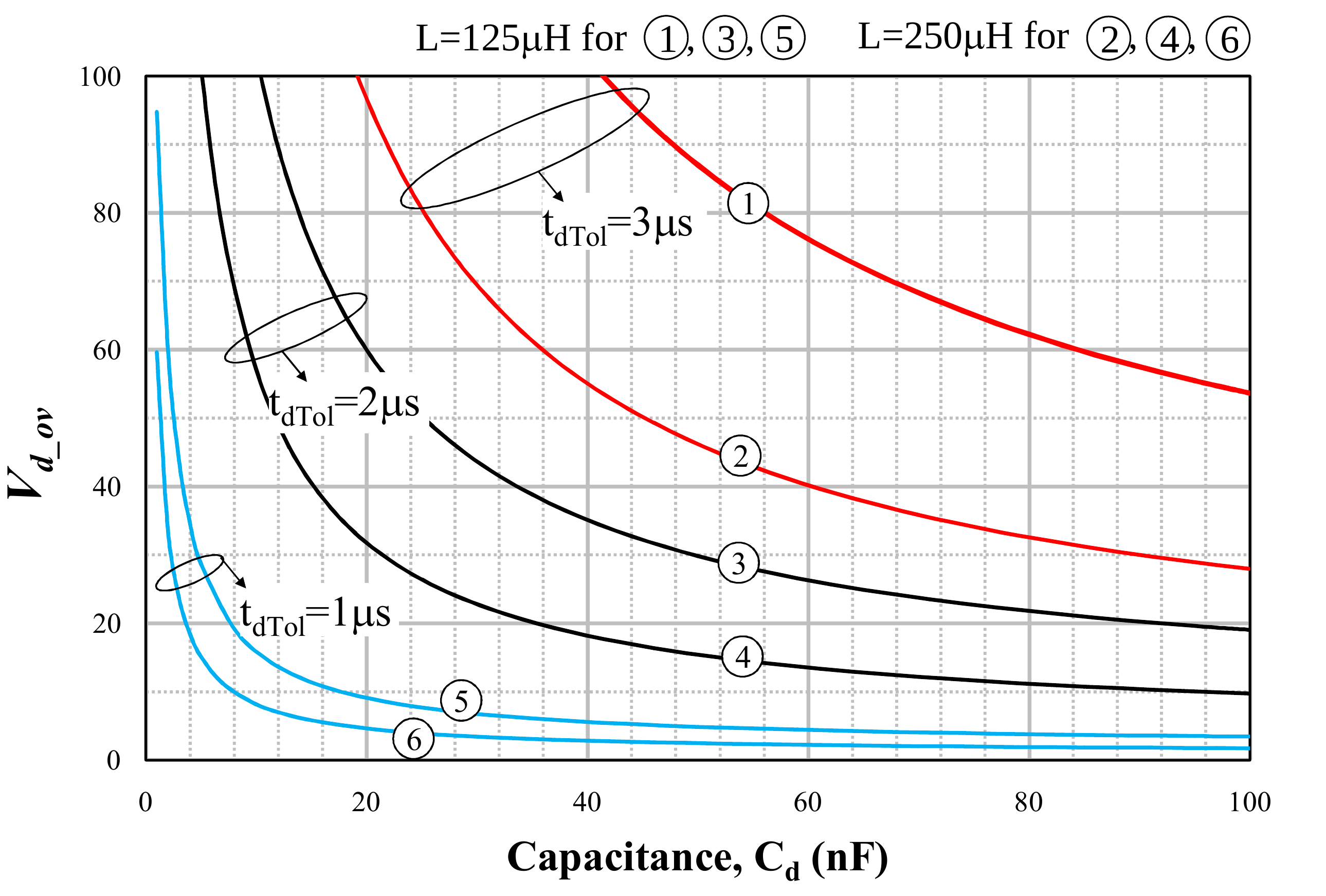}}\hspace{0.05in}
\subfigure[]{\includegraphics[keepaspectratio,scale=0.30]{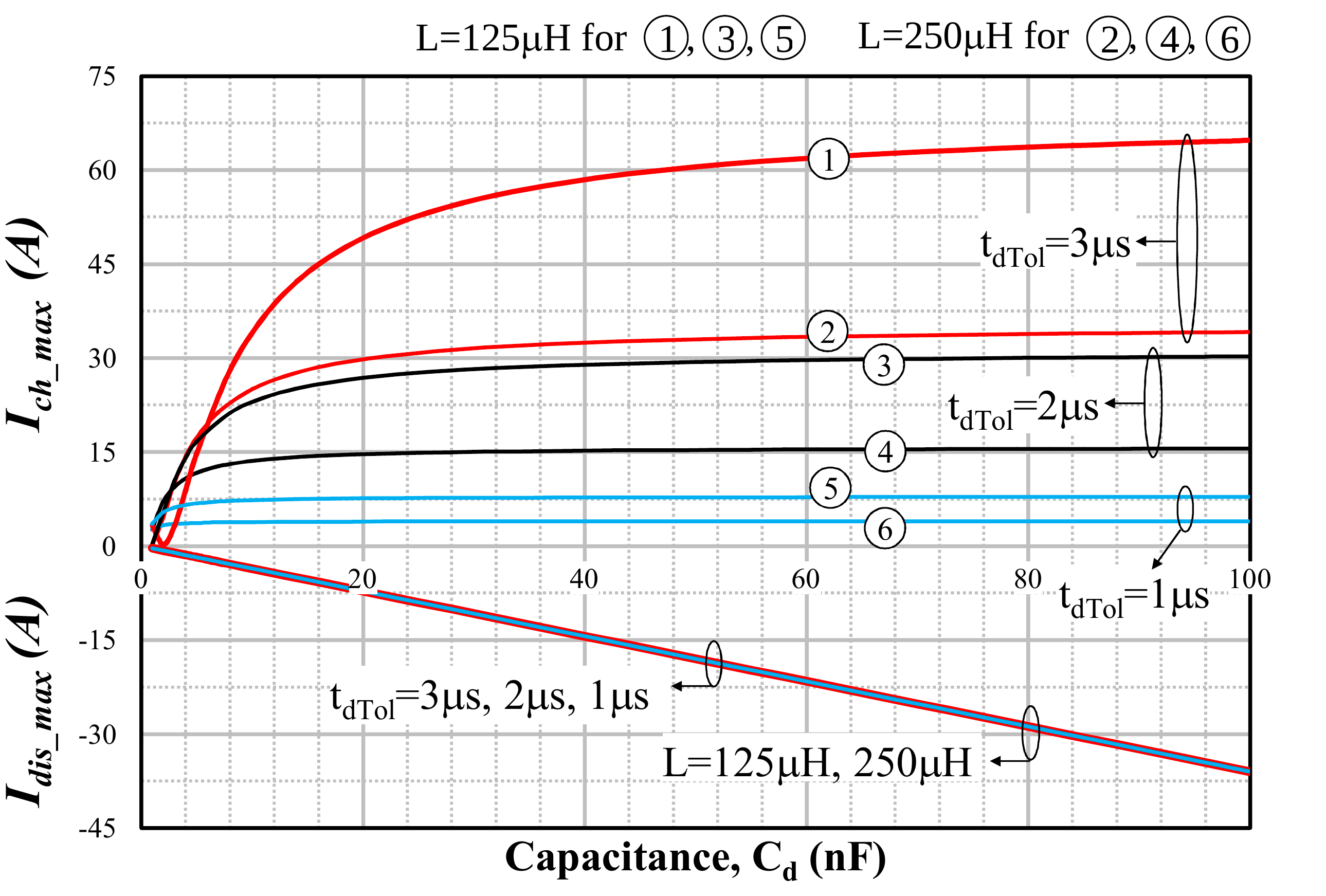}}
\caption{Variation in the circuit performance factors with $C_d$, for different $t_{dTol}$ and $L$, keeping $R_d=5\Omega$ and $t_{on}=5\mu s$
\subcaption{a.}{Percentage of over voltage $V_{d\_ov}$.}
\subcaption{b.}{$I_{ch\_max}$ (positive polarity) and $I_{dis\_max}$ (negative polarity).}}
\label{fig_di_dt}
\end{figure}
Since many practical thyristor characteristics have $t_{dTol}\le t_{on}$, this section analyses this case in detail. For the analysis consider a crowbar of $12kV$ rating built with $6$ thyristor devices in series and a $di/dt$ limiting inductor of $250\mu H$. The percentage over voltage appearing across the thyristor $T_1$ is defined as,
\begin{align}\label{overVolt}
V_{d\_ov}=\frac{V_{d1\_max}-(V_s/N)}{(V_s/N)}100
\end{align}
For various values of $C_d$, the $V_{d\_ov}$, $I_{ch\_max}$ and $I_{dis\_max}$ are computed from (\ref{overVolt}), (\ref{i1m1i1}) and (\ref{i3tmax}) respectively where $V_{d1\_max}$ is evaluated using (\ref{vd1m1i1}). The computation is carried out by choosing $t_{on}=5\mu s$ and keeping $t_{dTol}$ of thyristor and $R_{d}$ as parameters. The curve $V_{d\_ov}$ \textit{versus} $C_d$ is plotted in Fig.~\ref{fig_Vcurve}(a) for three different $t_{dTol}$ such as $3\mu s$, $2\mu s$ and $1\mu s$ and $R_{d}$ of $0\Omega$, $15\Omega$ and $30\Omega$. Fig.~\ref{fig_Vcurve}(a) shows that in all cases, beyond certain value of $C_d$ there is no significant reduction in $V_{d\_ov}$. Also from Figs.~\ref{fig_Vcurve}(b) and (c) higher value of $C_d$ leads to higher charge $I_{ch\_max}$ and discharge $I_{dis\_max}$ current. But the influence of $C_d$ on discharge current is found to be significant compared to its influence on charging current. Hence by choosing the minimum required value of $C_d$ that is sufficient to meet the required $V_{d\_ov}$, the discharge current can be limited to the minimum value. Fig.~\ref{fig_Vcurve}(a) also shows an increase in $V_{d\_ov}$ with resistance $R_d$ whereas charge and discharge currents reduces with $R_d$ for a given $C_d$. Hence by selecting a minimum value of $R_d$ the $V_{d\_ov}$ can be further reduced provided the charge and discharge currents are within the rating of $C_d$ as well as thyristor. For a given $C_d$ the $V_{d\_ov}$ also depends on $t_{dTol}$ shown in Fig.~\ref{fig_Vcurve}(a). For higher $t_{dTol}$ the $V_{d\_ov}$ is higher. Similar characteristics are observed for charging and discharging current, which are shown in Figs.~\ref{fig_Vcurve}(b) and (c) respectively. However the influence of $t_{dTol}$ on discharge current are found to be minimal for smaller values of $R_d$. Figs.~\ref{fig_VIton}(a), (b) and (c) shows the influence of $t_{on}$ on $V_{d\_ov}$, $I_{ch\_max}$ and $I_{dis\_max}$ for $R_d=5\Omega$. These curves are plotted for three different $t_{dTol}$ such as $3\mu s$, $2\mu s$ and $1\mu s$ and $t_{on}$ of $5\mu s$, $7.5\mu s$ and $10\mu s$. The $V_{d\_ov}$, $I_{ch\_max}$ and $I_{dis\_max}$ are highly influenced by $t_{on}$ and increases when $t_{on}$ reduces for a given $C_d$. The rate of increase in $V_{d\_ov}$, $I_{ch\_max}$ and $I_{dis\_max}$ for different $t_{on}$  is more when the ratio between $t_{on}$ and $t_{dTol}$ is less.

$I_{ratio}=I_{ch\_max}/I_{dis\_max}$ for different $C_d$ and $R_d=5\Omega$ is shown in Fig.~\ref{fig_VIton}(d). The curves are plotted for three different $t_{dTol}$ such as $3\mu s$, $2\mu s$ and $1\mu s$ and $t_{on}$ of $5\mu s$, $7.5\mu s$ and $10\mu s$. From Fig.~\ref{fig_VIton}(d) it can be observed that $I_{ratio}$ is approximately independent of $t_{on}$. Also for lower $C_d$ from Fig.~\ref{fig_VIton}(d) the charging current is higher than the discharging current. Hence for the design of dynamic balancing network $C_d$, it is necessary to compute both charging or discharging current.

The influence of $L$ on $V_{d\_ov}$, $I_{ch\_max}$ and $I_{dis\_max}$ are shown in Figs.~\ref{fig_di_dt}(a) and (b) respectively. For a given $C_d$ as $L$ increases $V_{d\_ov}$ and $I_{ch\_max}$ reduces whereas $I_{dis\_max}$ is independent of $L$. Hence for high $di/dt$ applications charging current can be higher than the discharging current. This shows the importance of dynamic balancing network when operating the crowbar at high $di/dt$ and importance of considering both currents for its design.
\subsection{Selection of dynamic balancing network ($R_d$,$C_d$)}
From the analysis for a given set of conditions $V_{d\_ov}$ shows minimum value for $R_d=0$. Hence to start, $C_d$ is computed for the maximum allowable transient voltage $V_{d1\_max}$ from (\ref{vd1m1i1}) keeping $R_d=0$. From the computed $C_d$ the $I_{ch\_max}$ and $I_{dis\_max}$ are calculated for $R_d=0$ from (\ref{i1m1i1}) and (\ref{i3tmax}) respectively. If these currents are within the acceptable limit then dynamic balancing network will have only $C_d$ of the computed value. If either $I_{ch\_max}$ or $I_{dis\_max}$ is higher than the specified value then from Figs.~\ref{fig_Vcurve}(b) and (c) the choice is to either increase $R_d$ or reduce $C_d$ by compromising on the required $V_{d\_ov}$. If significant increase of $R_d$ is required to bring $I_{ch\_max}$ and $I_{dis\_max}$ within limit, then reducing $C_d$ will be preferable approach. The other parameters $V_s$, $N$ and $L$ required for the computation are known from the crowbar circuit and $t_{dmax}$, $t_{dmin}$ and $t_{on}$ can be obtained from the selected thyristor datasheet.
\begin{table}[!t]
\processtable{Parameters of related to crowbar and thyristors\label{table_parameter}}
{\tabcolsep24pt
\begin{tabular}{|l|r|}
\hline
\bfseries Parameters & \bfseries Values\\
\hline
DC source voltage, $V_s$ & $12kV$\\
\hline
Series inductance to crowbar, $L$ & $250\mu H$\\
\hline
Rated dc voltage of thyristor, $V_{D(dc)}$ & $3.3kV$\\
\hline
Number of thyristor in series, $N$	& 6\\
\hline
Max. forward leakage current, $I_{Dmax}$	& $350\mu A$\\
\hline
Min. forward leakage current, $I_{Dmin}$ &	$100\mu A$\\
\hline
Max. turn ON delay time, $t_{dmax}$ &	$3\mu s$\\
\hline
Min. turn ON delay time, $t_{dmin}$ &	$0\mu s$\\
\hline
R.M.S. on state current, $I_{T(rms)}$ &	$550A$\\
\hline
Peak non-repetitive surge current, $I_{TSM}$ & $4500A$\\
\hline
Minimum recovery charge, $Q_{min}$ & $1000\mu C$\\
\hline
Maximum recovery charge, $Q_{max}$ & $2300\mu C$\\
\hline
Tolerance limit of capacitor, $a_C$ &	$0.1$\\
\hline
Tolerance limit of resistor, $a_R$ &	$0.05$\\
\hline
\end{tabular}}{}
\end{table}
\section{Simulation results}\label{Cdlimit}
From the system specification, the dc voltage rating of crowbar is 12kV. By connecting 6 thyristors having part number 5STP 03X6500 (ABB) in series the above voltage rating can be achieved. In nominal operating condition, each thyristor will see a voltage of $2kV$. Other important crowbar design parameters for this thyristor obtained from datasheet are given in Table~\ref{table_parameter}.
\subsection{Estimation of dynamic balancing capacitor ($C_d)$}
Choose the maximum allowable over voltage across the thyristor as $50\%$ ($1kV$) for a maximum value of $t_{on}$ equal to $5\mu s$ and $t_{dTol}$ of $3\mu s$. Then from Fig.~\ref{fig_Vcurve}(a) for $R_d=0$, the minimum value of $C_d$ required to limit $V_{d\_ov}$ with $50\%$ is $40nF$. Also from Fig.~\ref{fig_Vcurve}(a) with $C_d$ of $40nF$, $V_{d\_ov}$ will be less than $50\%$ if $t_{on}$ is higher than $5\mu s$ and $t_{dTol}$ less than $3\mu s$. Since the $C_d$ experience $3kV$ the voltage rating of $C_d$ is chosen as $4kV$.
\subsection{Estimation of dynamic balancing resistor ($R_d$)}\label{s52}
The $I_{ch\_max}$ and $I_{dis\_max}$ for $t_{on}=5\mu s$, $t_{dTol}=3\mu s$ and $R_d=0$ is recorded from Figs.~\ref{fig_Vcurve}(b) and (c) respectively. From Fig.~\ref{fig_Vcurve}(b) the $I_{ch\_max}$ for $C_d=40nF$ is $33A$ whereas from Fig.~\ref{fig_Vcurve}(c) the discharging current $15A$. Since these current are small compared to the current rating of thyristor, given in Table~\ref{table_parameter}, $C_d$, can be directly connected without any current limiting resistor. However a small damping resistance of $3\Omega$ is chosen for $R_d$. This is sufficient considering a parasitic loop inductance of the $(R_d,C_d)$ balancing network and thyristor is of the order of $100nH$.
\begin{figure*}[!t]
  \centering
\subfigure[]{\includegraphics[keepaspectratio,scale=0.33]{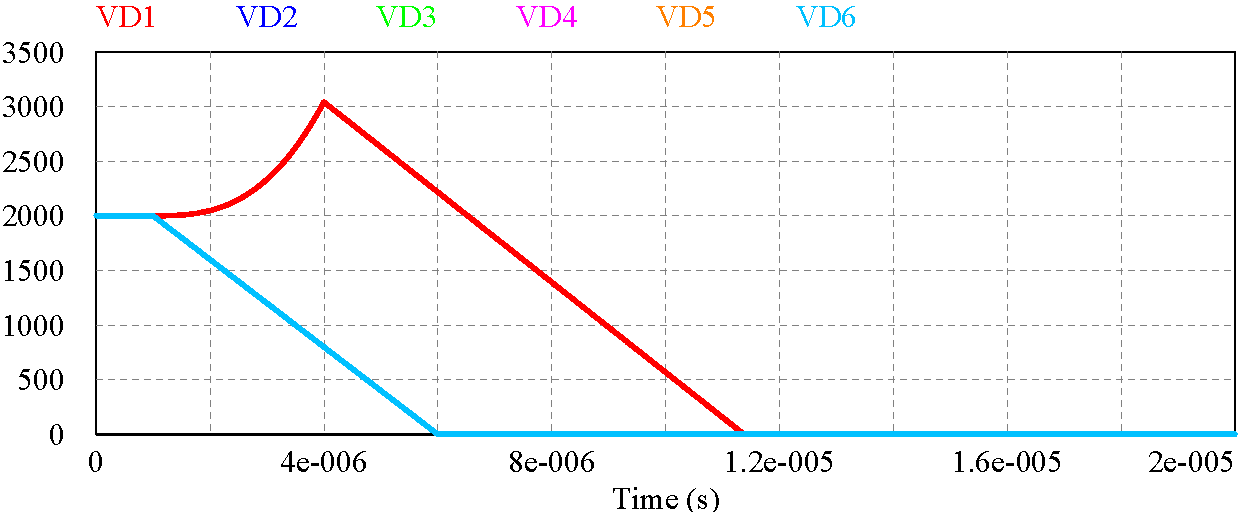}}\hspace{0.15in}
\subfigure[]{\includegraphics[keepaspectratio,scale=0.40]{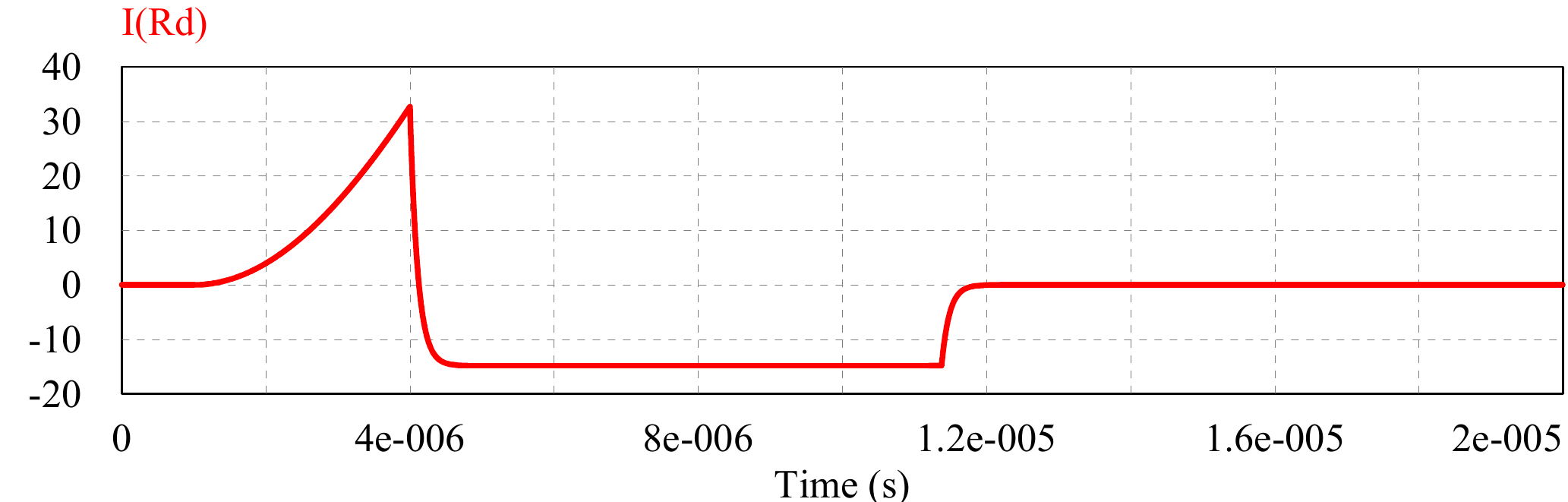}}
\small
\put(-389,60){\textit{VD1}}
\put(-412,38){\textit{VD2 to VD6}}
\put(-210,45){\textit{$i_{ch}(t)$}}
\put(-150,25){\textit{$i_{dis}(t)$}}
\put(-460,30){\rotatebox{90}{$v_{AK}$\textit{(V)}}}
\put(-244,36){\rotatebox{90}{$(A)$}}\\
\hspace{0.10969in}
\subfigure[]{\includegraphics[keepaspectratio,scale=0.60]{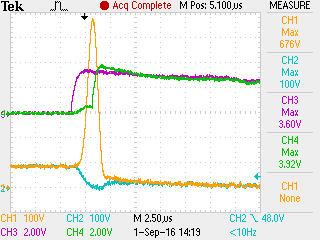}}
\put(-172,85){\textsl{\small V\raisebox{-.4ex}{\scriptsize G,T\raisebox{-.4ex}{\tiny 2}}}}
\put(-135,85){\textsl{\small V\raisebox{-.4ex}{\scriptsize G,T\raisebox{-.4ex}{\tiny 1}}}}
\put(-131,55){\textsl{\small V\raisebox{-.4ex}{\scriptsize AK,T\raisebox{-.4ex}{\tiny 1}}}}
\put(-170,32){\textsl{\small V\raisebox{-.4ex}{\scriptsize AK,T\raisebox{-.4ex}{\tiny 2}}}}
\put(-90,130){\textsl{\scriptsize $V_{G}:$~2V/div.}}
\put(-90,122){\textsl{\scriptsize $V_{AK}:$~100V/div.}}
\put(-173,126){\textsl{\scriptsize $2.5\mu s$}}
\put(-166.2,122){\vector(1,0){10}}
\put(-161.5,122){\vector(-1,0){10}}
\hspace{0.6in}
\subfigure[]{\includegraphics[keepaspectratio,scale=0.60]{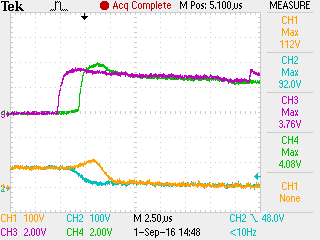}}
\put(-180,85){\textsl{\small V\raisebox{-.4ex}{\scriptsize G,T\raisebox{-.4ex}{\tiny 2}}}}
\put(-140,85){\textsl{\small V\raisebox{-.4ex}{\scriptsize G,T\raisebox{-.4ex}{\tiny 1}}}}
\put(-130,42){\textsl{\small V\raisebox{-.4ex}{\scriptsize AK,T\raisebox{-.4ex}{\tiny 1}}}}
\put(-170,32){\textsl{\small V\raisebox{-.4ex}{\scriptsize AK,T\raisebox{-.4ex}{\tiny 2}}}}
\put(-90,130){\textsl{\scriptsize $V_{G}:$~2V/div.}}
\put(-90,122){\textsl{\scriptsize $V_{AK}:$~100V/div.}}
\put(-173,126){\textsl{\scriptsize $2.5\mu s$}}
\put(-166.2,122){\vector(1,0){10}}
\put(-161.5,122){\vector(-1,0){10}}
\hspace{0.3in}
   \caption{Simulation waveforms (a. \& b.) and experimental waveforms (c. \& d.) considering worst case delay
   \subcaption{a.}{Simulation results showing voltage across all 6 thyristors, considering worst case delay and parameter tolerance for the first thyristor.}
   \subcaption{b.}{Simulation results showing charging $i_{ch}(t)$ and discharging $i_{dis}(t)$ current of the first thyristor.}
   \subcaption{c.}{For $t_{dTol}=2400ns$, $v_{AK}$ waveforms across $T_1$ and $T_2$ with respective gate signals without balancing network.}
   \subcaption{d.}{For $t_{dTol}=2400ns$, $v_{AK}$ waveforms across $T_1$ and $T_2$ with respective gate signals with balancing network.}} 
   \label{fig_sim1}
\end{figure*}
\begin{figure}[!t]
\centering
\subfigure[]{\includegraphics[keepaspectratio,scale=0.60]{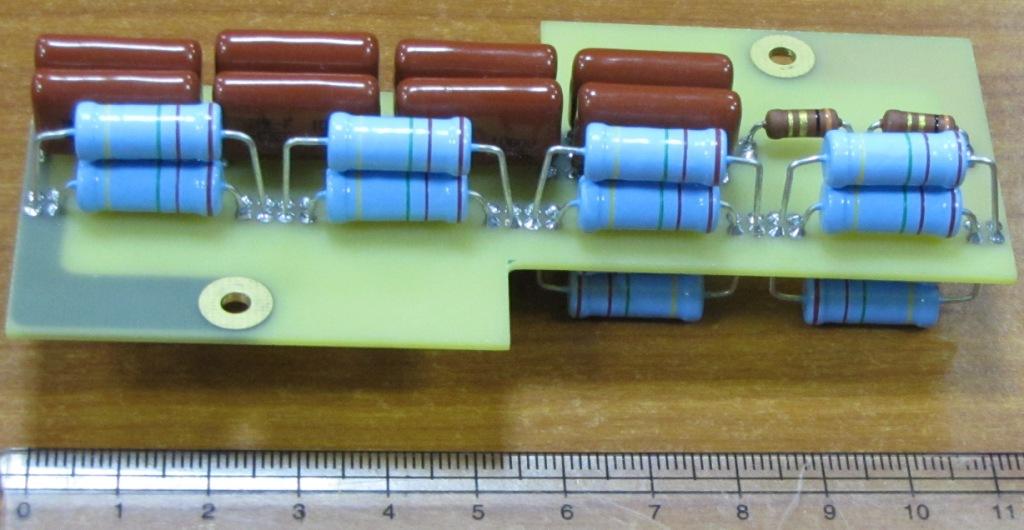}}
\put(-100,114){\textsl{\Large (1)}}
\put(-60,106){\textsl{\Large (2)}}
\put(-40,83){\textsl{\Large (3)}}\\
\subfigure[]{\includegraphics[keepaspectratio,scale=0.60]{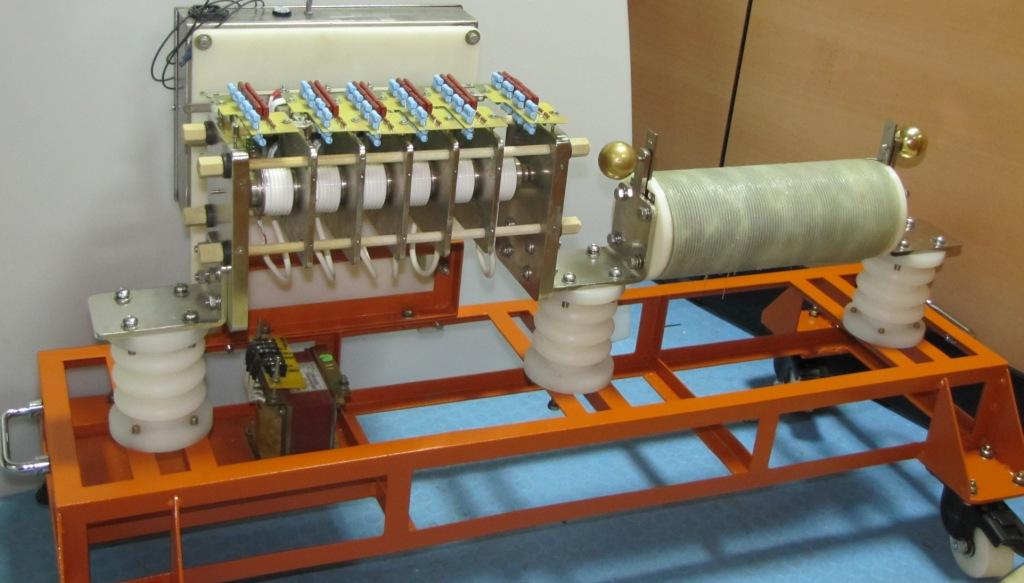}}
\put(-60,85){\textsl{\Large $L$}}
\put(-200,125){\textsl{\large Thyristor stack}}\\
\subfigure[]{\includegraphics[keepaspectratio,scale=0.60]{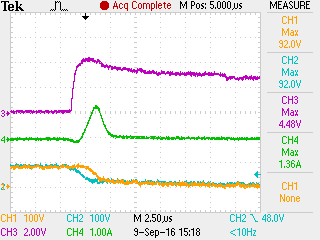}}
\put(-170,100){\textsl{\small $v$\raisebox{-.4ex}{\scriptsize G,T\raisebox{-.4ex}{\tiny 1}}}}
\put(-130,78){\textsl{\small $i$\raisebox{-.4ex}{\scriptsize dis}(t)}}
\put(-153,52){\textsl{\small $i$\raisebox{-.4ex}{\scriptsize ch}(t)}}
\put(-135,42){\textsl{\small $v$\raisebox{-.4ex}{\scriptsize AK,1}}}
\put(-165,33){\textsl{\small $v$\raisebox{-.4ex}{\scriptsize AK,2}}}
\put(-90,128){\textsl{\scriptsize $v_{G}:$~2V/div.}}
\put(-90,120){\textsl{\scriptsize $v_{AK}:$~100V/div.}}
\put(-90,112){\textsl{\scriptsize $i_{ch}:$~1A/div.}}
\put(-173,126){\textsl{\scriptsize $2.5\mu s$}}
\put(-166.2,122){\vector(1,0){10}}
\put(-161.5,122){\vector(-1,0){10}}
\hspace{0.3in}
\subfigure[]{\includegraphics[keepaspectratio,scale=0.32]{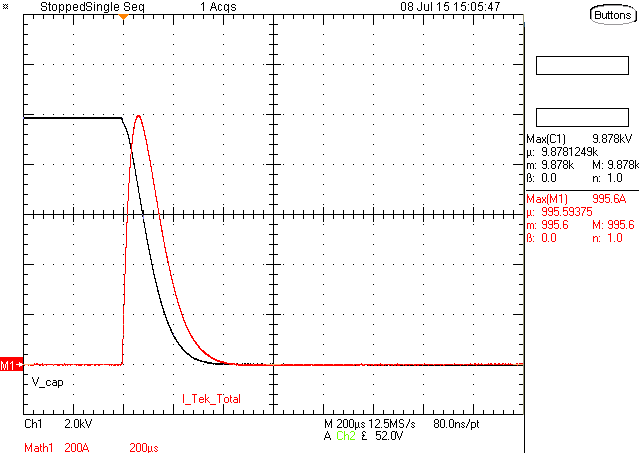}}
\put(-193,126){\scriptsize Crowbar}
\put(-196,119){\scriptsize voltage,V\raisebox{-.4ex}{\scriptsize s}}
\put(-195.4,111){\scriptsize (2kV/div.)}
\put(-155,95){{\scriptsize Crowbar current (200A/div.)}}
\put(-71.5,129){\textsl{\scriptsize $200\mu s$}}
\put(-63.2,125){\vector(1,0){10}}
\put(-59.5,125){\vector(-1,0){10}}
\caption{Showing
\subcaption{a.}{Photographs of static and dynamic balancing network where the component (1) is $C_d$ (2) is $R_d$ and (3) is $R_s$.}
\subcaption{b.}{Photographs of $12kV$, $1kA$ solid state crowbar.}
\subcaption{c.}{Charging (negative polarity) and discharging (positive polarity) current from dynamic balancing network connected to $T_1$ along with its gate signal and voltage waveform across $T_1$ and $T_2$.}
\subcaption{d.}{Nominal voltage $V_s$ ($10kV$) applied across crowbar and current ($1kA_{peak}$) through crowbar.}}
\label{fig_sim_}
\end{figure}
\subsection{Estimation of static balancing resistor ($R_s$)}
Based on the parameter given in Table~\ref{table_parameter} and limiting the maximum steady state voltage across any of the thyristor $V_{d1}$ to $135\%$ ($2.7kV$), the $R_s$ can be estimated from (\ref{Rseq}) as $2.5M\Omega$.

Simulations are carried out with estimated values of static and dynamic balancing network elements. In simulation $t_{on}$ and $t_{dTol}$ is chosen as $5\mu s$ and $3\mu s$ respectively. The voltage waveform across each thyristor during turn-on is shown in Fig.~\ref{fig_sim1}(a). The charging and discharging current of $C_d$ are shown by positive and negative polarity respectively in Fig.~\ref{fig_sim1}(b). Simulation results shows that the maximum dynamic voltage reaches $3kV$ and matches the target design of $V_{d\_ov}$ of 50\% in the analytical design. Also from simulation the $I_{ch\_max}$ and $I_{dis\_max}$ through $C_d$ is found to be $33A$ and $15A$ respectively that also matches with the analytical design performance from section~\ref{s52}.
\section{Comparison with dynamic balancing network based on reverse recovery charge}
Design of $C_d$ based on reverse recovery charge is reported in~\cite{IEEEhowto:link1}. Value of $C_d$ based on reverse recovery charge is,
\begin{align}
C_d=\frac{\left(1+(N-1)\dfrac{1-a_C}{1+a_C}\right)(Q_{max}-Q_{min})}{(1-a_C)\left[V_{d1}\left(1+(N-1)\dfrac{1-a_C}{1+a_C}\right)-V_s\right]}
\end{align}

Using data given in Table~\ref{table_parameter} and limiting the same maximum allowable over voltage across thyristor to 50\%, the value of $C_d$ based on reverse recovery charge is computed as $2.25\mu F$.

From (\ref{i1m1i1}) and (\ref{i3tmax}) the $I_{ch\_max}$ and $I_{dis\_max}$ for the above $C_d$ with $t_{dTol}=3\mu s$ and $t_{on}=5\mu s$ is $36A$ and $810A$ respectively. Comparison with the values obtained from proposed design of dynamic balancing network, the $I_{ch\_max}$ is comparable where as the $I_{dis\_max}$ is significantly large, closer to the rating of thyristor given in Table~\ref{table_parameter}. To reduce the discharging current $R_d$ of large value of resistance and wattage is required.

Comparing the values of $C_d$ based on turn-on delay time and based on reverse recovery charge, it is found that $C_d$ with reverse recovery charge is $56$ times larger than that of $C_d$ obtained with turn-on delay time. By Fig.~\ref{fig_Vcurve}(a) such a large value of $C_d$ will not give any additional reduction of $V_{d\_ov}$, instead it significantly increases the discharging current. This large value of discharging current as well as large value of capacitance makes the dynamic balancing network lossy, bulky as well as costly.
\section{Experimental Results}
\begin{figure*}[!t]
  \centering
\subfigure[]{\includegraphics[keepaspectratio,scale=0.60]{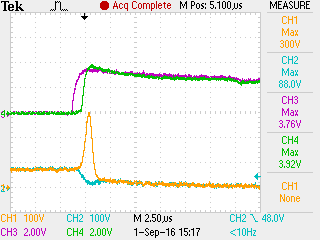}}
\put(-170,90){\textsl{\small V\raisebox{-.4ex}{\scriptsize G,T\raisebox{-.4ex}{\tiny 2}}}}
\put(-140,90){\textsl{\small V\raisebox{-.4ex}{\scriptsize G,T\raisebox{-.4ex}{\tiny 1}}}}
\put(-135,65){\textsl{\small V\raisebox{-.4ex}{\scriptsize AK,T\raisebox{-.4ex}{\tiny 1}}}}
\put(-170,32){\textsl{\small V\raisebox{-.4ex}{\scriptsize AK,T\raisebox{-.4ex}{\tiny 2}}}}
\put(-90,130){\textsl{\scriptsize $V_{G}:$~2V/div.}}
\put(-90,122){\textsl{\scriptsize $V_{AK}:$~100V/div.}}
\hspace{0.3in}
\subfigure[]{\includegraphics[keepaspectratio,scale=0.60]{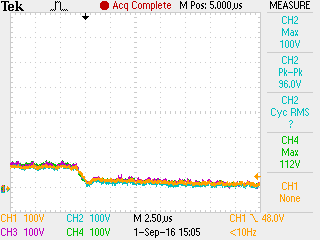}}
\put(-180,53){\textsl{\small V\raisebox{-.4ex}{\scriptsize AK,T\raisebox{-.4ex}{\tiny 3}} to V\raisebox{-.4ex}{\scriptsize AK,T\raisebox{-.4ex}{\tiny 6}}}}
\put(-90,130){\textsl{\scriptsize $V_{AK}:$~100V/div.}}
\hspace{0.3in}
\subfigure[]{\includegraphics[keepaspectratio,scale=0.60]{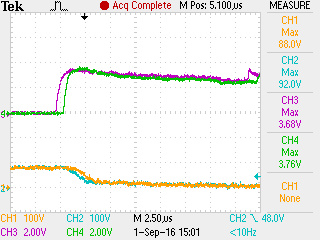}}
\put(-180,90){\textsl{\small V\raisebox{-.4ex}{\scriptsize G,T\raisebox{-.4ex}{\tiny 2}}}}
\put(-150,90){\textsl{\small V\raisebox{-.4ex}{\scriptsize G,T\raisebox{-.4ex}{\tiny 1}}}}
\put(-140,40){\textsl{\small V\raisebox{-.4ex}{\scriptsize AK,T\raisebox{-.4ex}{\tiny 1}}}}
\put(-170,32){\textsl{\small V\raisebox{-.4ex}{\scriptsize AK,T\raisebox{-.4ex}{\tiny 2}}}}
\put(-90,130){\textsl{\scriptsize $V_{G}:$~2V/div.}}
\put(-90,122){\textsl{\scriptsize $V_{AK}:$~100V/div.}}
\hspace{0.3in}
\subfigure[]{\includegraphics[keepaspectratio,scale=0.60]{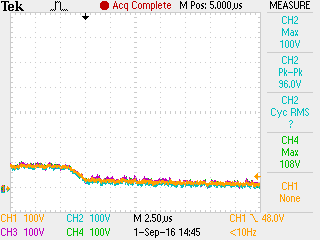}}
\put(-180,53){\textsl{\small V\raisebox{-.4ex}{\scriptsize AK,T\raisebox{-.4ex}{\tiny 3}} to V\raisebox{-.4ex}{\scriptsize AK,T\raisebox{-.4ex}{\tiny 6}}}}
\put(-90,130){\textsl{\scriptsize $V_{AK}:$~100V/div.}}
   \caption{For $t_{dTol}=800ns$, the $v_{AK}$ waveform across~~(Time scale: $2.5\mu s/div.$)
   \subcaption{a.}{$T_1$ and $T_2$ with respective gate signals without balancing network.}
   \subcaption{b.}{$T_3$ to $T_6$ without balancing network.}
   \subcaption{c.}{$T_1$ and $T_2$ with respective gate signals with balancing network.}
   \subcaption{d.}{$T_3$ to $T_6$ with balancing network}.} 
   \label{fig_sim}
\end{figure*}

Static and dynamic balancing network are fabricated for a crowbar of $12kV$, $1kA$ rating. 
The dynamic balancing components $C_d$ and $R_d$ chosen for the experiment are $47nF$ and $3\Omega$ respectively. The value of static balancing resistance $R_s$ used is $2.2M\Omega$. Both static and dynamic balancing network required for one thyristor are assembled on a single four layer PCB, where interconnecting tracks are only routed through inner layers. The top and bottom layer provides isolation to the inner tracks from external media. The assembled static and dynamic balancing network is shown in Fig.~\ref{fig_sim_}(a). These balancing network PCBs are mounted on crowbar unit as shown in Fig.~\ref{fig_sim_}(b). The performance of balancing circuit is evaluated in lower voltage by choosing $6$ numbers of thyristors $T_1$ to $T_6$ having similar $t_{dTol}$. A known delay is introduced to the gate signal of first thyristor $T_1$ and the voltages across all six thyristors are recorded with and without the voltage balancing network. The experiment is carried out with a dc voltage of $480V$ and $t_{dTol}$ of $800ns$ and $2400ns$. The $t_{dTol}$ of $800ns$ emulate the difference in propagation delay among gate signals of thyristors and $2400ns$ is typically the maximum difference in turn-on delay time among thyristors. The $t_{on}$ observed for the thyristors used for the experiment is $3\mu s$. The other parameters are as given in Table~\ref{table_parameter}. The voltage waveforms across the thyristors during turn-on without balancing network for $t_{dTol}=800ns$ are shown in Figs.~\ref{fig_sim}(a) and (b) where Fig.~\ref{fig_sim}(a) are for $T_1$ and $T_2$ and Fig.~\ref{fig_sim}(b) are for $T_3$ to $T_6$. An appreciable over voltage of $3.75$ times its steady state voltage is observed across the delayed thyristor $T_1$ even for small value of $t_{dTol}$. The effectiveness of balancing network is shown in Figs.~\ref{fig_sim}(c) and (d) where Fig.~\ref{fig_sim}(c) are for $T_1$ and $T_2$ and Fig.~\ref{fig_sim}(d) are for $T_3$, $T_4$, $T_5$ and $T_6$. The over voltage across $T_1$ with balancing network is found to be insignificant as shown in Fig.~\ref{fig_sim}(c). The over voltage computed analytically with (\ref{vd1m1i1}) is $82V$ which is $2\%$ above its steady state value that is close to the experimental result shown in Fig.~\ref{fig_sim}(c). The charging and discharging current of dynamic balancing network is shown in Fig.~\ref{fig_sim_}(c) measured to be $0.2A$ and $-1.2A$ respectively where are the respective computed values from (\ref{i1m1i1}) and (\ref{i3tmax}) are $0.18A$ and $-1.14A$ closely matching with the experimental results.

The experiment is repeated with $t_{dTol}=2400ns$ that is close to the maximum value mentioned in the thyristor datasheet~\cite{IEEEhowto:abb}. The voltage waveforms across $T_1$ and $T_2$ during turn-on without balancing network are shown in Fig.~\ref{fig_sim1}(c) where as that of $T_3$ to $T_6$ are not shown since they are triggered simultaneously. The over voltage across $T_1$ is found to be $676V$ that is $8.45$ times that of the steady state value of $80V$. The voltage waveforms across $T_1$ and $T_2$ with balancing network is shown in Fig.~\ref{fig_sim1}(d) where the over voltage across $T_1$ is measured to be $112V$ that is $1.4$ times of its steady state value. The computed over voltage from (\ref{vd1m1i1}) is found to be $112.6V$ which is close to the experimental result and percentage of over voltage is close to the simulation results shown in Fig.~\ref{fig_sim1}(a). The crowbar is operated by discharging the nominal forward blocking voltage of $10kV$ to allow nominal current of $1kA$. The applied forward voltage and crowbar current are shown in Fig.~\ref{fig_sim_}(d). The performance of balancing network and crowbar are found to be satisfactory and the proposed design procedure for the balancing network is validated with analytical, simulation and experimental results.
\section{Conclusion}
Voltage balancing networks are often designed by considering two modes of operation of thyristors which are reverse blocking mode and turn-off mode. Since in a crowbar application, the modes of operation of thyristor are different, the conventional method used for the design of balancing network if adopted, leads to very bulky balancing components with higher power loss. This paper proposes a design method for dynamic balancing network based on gate turn-on delay time. The paper derives two models for the dynamic balancing network and shows its importance in the design of dynamic balancing network when crowbar operate at high $di/dt$. The proposed approach for designing that balancing network results in a  small value of capacitance as well as a small value of discharging current which makes the dynamic balancing network more efficient and compact. The influence of dynamic balancing resistance and crowbar current limiting inductance on voltage imbalance, charging current and discharging current is explained. This method also allows one to operate a series connected string of thyristors without any complex pulse synchronizing circuit that is normally found in crowbar applications. The analysis done for the balancing network includes component tolerance to capture the worst case circuit operating conditions. For the experimental validation two practically encountered delays, the difference in propagation delay among gate signals of thyristors and difference in turn-on delay among thyristors, are considered. Experimental results on a $12kV$, $1kA$ crowbar shows excellent results and confirms the theoretical analysis and the proposed design procedure.
\section{Acknowledgment}
The work is supported by Ministry of Electronics and Information Technology, Govt. of India, through NaMPET programme and Department of Atomic Energy (DAE), Government of India through Institute for Plasma Research, Gandhinagar, India. The authors thank Mr. Rajiv I. at C-DAC, Thiruvananthapuram for support with the experimental measurements.

\bibliographystyle{IEEEtran}
\bibliography{IEEEabrv,References}
%

\end{document}